\documentclass{revtex4-1}
\usepackage[applemac]{inputenc}
\usepackage{amsmath,amssymb}
\usepackage{graphicx}
\usepackage{float}
\usepackage{hyperref}
\begin{document}
\title{THE BONDI PROBLEM REVISITED: \\A SPECTRAL DOMAIN DECOMPOSITION CODE}
\author{M. A. ALCOFORADO}
\address{Departamento de F\'{\i}sica Te\'orica - Instituto de F\'{\i}sica
	A. D. Tavares, Universidade do Estado do Rio de Janeiro,
	R. S\~ao Francisco Xavier, 524. Rio de Janeiro, RJ, 20550-013, Brazil
malcoforado@hotmail.com}
\author{W. O. BARRETO}
\address{Centro de Ci\^encias Naturais e Humanas, Universidade Federal do ABC,\\Av. dos Estados 5001, 09210-580 Santo Andr\'e, S\~ao Paulo, Brazil\\willians.barreto@ufabc.edu.br\\
Centro de F\'{\i}sica Fundamental, Universidad de Los Andes, M\'erida 5101,  Venezuela}
\author{H. P. DE OLIVEIRA}
\address{Departamento de F\'{\i}sica Te\'orica - Instituto de F\'{\i}sica
	A. D. Tavares, Universidade do Estado do Rio de Janeiro, 
	R. S\~ao Francisco Xavier, 524. Rio de Janeiro, RJ, 20550-013, Brazil\\
henrique.oliveira@uerj.br}
\begin{abstract}
We present a simple domain decomposition code based on the Galerkin-Collocation method to integrate the field equations of the Bondi problem. The algorithm is stable, exhibits exponential convergence when considering the Bondi formula as an error measure, and is computationally economical. We have incorporated features of both Galerkin and Collocation methods along with the establishment of two non-overlapping subdomains. We have further applied the code to show the decay of the Bondi mass in the nonlinear regime and its power-law late time decay. Another application is the determination of the wave-forms at the future null infinity connected with distinct initial data.
\end{abstract}
\maketitle
%%%%%%%%%%%%%%%%%%%%%%
\section{Introduction}
%%%%%%%%%%%%%%%%%%%%%%

The recent direct detection of gravitational waves by the LIGO consortium \cite{LIGO_gws} opened a new window to observe the universe. Gravitational waves have a unique feature of extracting mass from the source besides carrying relevant information about their origin. The observed signals were identified as produced by a collision of spinning black holes after comparing them with those waveforms generated by successful long-term simulations of binary black holes \cite{pretorius,campanelli,baker,mroue,husa}. For this reason, numerical relativity has become a relevant and mature area of investigation where we can uncover the consequences of the gravitational strong field regime. 

A crucial step of understanding the emission of gravitational waves by an isolated source was provided by the seminal paper of Bondi and collaborators \cite{bondi}. The decay of the Bondi mass as a consequence of the emission of gravitational waves in connection with the news function is the central result that is summarized in the Bondi formula. With Sachs \cite{sachs} and Penrose's works \cite{penrose}, Bondi et al.'s contribution constitute the cornerstones of general relativity's characteristic formulation, particularly tailored to study gravitational radiation. The review of Winicour \cite{winicour_lrr} presents a self-contained and complete discussion on the characteristic formulation of the field equations. 

%Gomez et al. proposed the first numerical code to integrate the Bondi equations known as the PITT code.  

The first numerical code to evolve the Bondi field equations, and later extended to the more general Bondi-Sachs problem, is the PITT code \cite{papadopoulos}. In a series of papers, the code showed up to be stable, second-order accurate, and fully nonlinear, besides several applications in situations of physical interest \cite{isaacson,bishop,babiuc}. Another feature of the PITT code is the calculation of the waveforms at null infinity \cite{bishop_97,babiuc_2011_1,babiuc_2011_2}.

We have proposed the first code based on the Galerkin and Collocation methods to evolve the Bondi equations several years ago \cite{rodrigues}, later Handmer and Szilagyi implemented a spectral code for the Bondi-Sachs equations \cite{handmer}. We highlight two features of the spectral algorithm: the low computational cost to achieve good accuracy and combined aspects of the Galerkin and Collocation methods. For the sake of clarifying this last aspect, we provide below a brief description of the basic structure we have employed in the present formulation. 

Let us consider a function $f(x,t)$ that satisfies a differential equation, say $f_{,t}-L(f,f_{,x},..)=0$ in a spatial domain $a \leq x \leq b$, where $L$ is a nonlinear function. Spectral methods belong to a general class of the weighted residual methods (WRM) \cite{finlayson} that establish an alternative strategy to solve any differential equation. The first step is to approximate the function $f(x,t)$ by a finite series expansion with respect to a set of analytical functions (for instance, Chebyshev or Legendre polynomials) known as the trial or basis functions. Accordingly, we have

\begin{equation}
f_N(x,t)=\sum_{k=0}^N\,\hat{f}_k(t) \psi_k(x), \label{eq1}
\end{equation}

\noindent where $\hat{f}_k(t)$ are the unknown coefficients or modes, $N$ is the truncation order that dictates the number of modes, and $\psi_k(x)$ represents the basis functions. We choose the modes $\hat{f}_k(t)$ such that an error measure provided by the residual equation $\mathrm{Res}_N(x,t) = f_{N,t} - L(f_N,f_{N,x},..)$ is forced to be zero in an average sense, meaning that the weighted integral of the residual equation are set to zero, or 

\begin{equation}
\left<\mathrm{Res}_N(x,t),\phi_j(x)\right> = \int_{a}^{b}\,\mathrm{Res}_N(x,t) \phi_j(x)w(x) dx = 0, \label{eq2}
\end{equation} 

\noindent for all $j=0,1,..,N$. Here $\phi_j(x)$ are the test functions and $w(x)$ is the associated weight. The choice of the test functions specifies the type of spectral method we are adopting \cite{boyd,canuto}. For instance, if $\phi_j(x)=\psi_k(x)$, and choosing the basis functions to satisfy the boundary conditions, we have the traditional Galerkin method. On the other hand, if $\phi_j(x) = \delta(x-x_j)$, where $x_j,\,j=0,1,..,N$ are the collocation points, we obtain the Collocation method. In this case, we can infer from Eq. (2) that the residual equation vanishes at each collocation point, $\mathrm{Res}_N(x_j,t),\,j=0,1,..,N$ (here $w=1$).  Consequently, any spectral method approximates an evolution partial differential equation as a finite set of ordinary differential equations for the modes $\hat{f}_k$ or the values of $f(x,t)$ at the collocation points, $f_j(t)=f_N(x_j,t)$ in the case of the Collocation method. For an elliptic-type equation, spectral methods approximate it as a finite set of algebraic equations related to the modes. 

We have combined features of the Galerkin and Collocation methods to adopt basis functions that satisfy the boundary conditions and assume that the test functions are Dirac functions. Further, in some cases, we evaluate the integrals (2) with quadrature formulae, a prescription of the Galerkin method with numerical integration (G-NI) \cite{canuto}. We have coined this combination by the Galerkin-Collocation method and applied it to several problems: the Bondi problem \cite{rodrigues}, the spherical collapse of scalar fields and critical phenomena \cite{rodrigues_spherical,crespo_bh_scalar_field,crespo_kink,crespo_affine,alcoforado_critical,alcoforado_cauchy}, the evolution of cylindrical waves \cite{celestino_cylindrical,barreto_cylindrical_DD}, and the initial data for numerical relativity \cite{matias_sd,barreto_DD,barreto_DD_2}. 

Recently, we have incorporated the technique of domain decomposition into the spectral Galerkin-Collocation scheme (see Refs. \cite{crespo_affine,alcoforado_critical,alcoforado_cauchy,barreto_cylindrical_DD,barreto_DD,barreto_DD_2}). The domain decomposition or multidomain method consists of dividing the spatial domain into two or more subdomains and establishing appropriate transmission conditions to connect the solutions in each subdomain. Domain decomposition is adequate to tackle complicated geometries or exhibit strong field gradients in some areas of the spatial domain. The use of this technique is not new in numerical relativity when combined with spectral methods \cite{bona,pfeiffer,ansorg,spec,lorene,szilagyi_09,kidder1,kidder2,hemberger_13,sxs_col}. 

The present paper aims to incorporate the domain decomposition technique in the Galerkin-Collocation code for the Bondi equations. In Section II, we present the metric, the field equations of the Bondi problem, and we discuss the evolution scheme provided by the characteristic scheme. We also show the boundary conditions imposed on the metric functions to satisfy the requirements of spacetime regularity and asymptotic flatness and the Bondi formula. We provide a detailed description of the domain decomposition algorithm in Section III. We have defined the basis functions in each subdomain, the transmission conditions, and the procedure to approximate the hypersurface and evolution equations into sets of finite algebraic and dynamical systems, respectively. Section IV is devoted to the code validation with the convergence tests using the Bondi formula.

Some physical applications are presented in Section V. We have exhibited the power-law Bondi mass decay for late times. We have also considered several examples of the decay of the Bondi mass for higher initial amplitudes and the gravitational waveforms generated in each case. Finally, in Section VI we conclude.

%%%%%%%%%%%%%%%%%%%%%%%%%%%%%%
\section{The field equations}%
%%%%%%%%%%%%%%%%%%%%%%%%%%%%%%

The metric established by Bondi, van der Burgh and Metzner \cite{bondi} that describes axisymmetric and asymptotic spacetimes takes the form

%\begin{widetext}
\begin{eqnarray}
ds^2=-\left(\frac{V}{r}\mathrm{e}^{2\beta}-U^2 r^2 \mathrm{e}^{2\gamma}\right) du^2 - 2\mathrm{e}^{2\beta} du dr \nonumber \\
\nonumber \\
- 2 U r^2 \mathrm{e}^{2\gamma} du d\theta + r^2(\mathrm{e}^{2\gamma} d \theta^2 + \mathrm{e}^{-2\gamma}\sin^2 \theta d\varphi^2). \label{eq3}
\end{eqnarray}
%\end{widetext}

\noindent Here, $u$ is the retarded time coordinate such that $u=constant$ denotes the outgoing null cones; the radial coordinate $r$ is chosen demanding that the surfaces $(u,r)$ have area equal to $4 \pi r^2$ and the angular coordinates $(\theta,\varphi)$ are constant alogn the outgoing null geodesics. The metric functions $\gamma, \beta, U$ and $V$ depend on the coordinates $u,r,\theta$ and satisfy the vacuum field equations $R_{\mu\nu} = 0$. Following Bondi et al \cite{bondi}, the equations are organized into two sets: three hypersurface equations and one evolution equations, given, respectively, by: 

%\begin{widetext}
\begin{eqnarray}
\beta_{,r} &=& \displaystyle{\frac{1}{2} r(\gamma_{,r})^2}, \label{eq4}\\
\nonumber \\
\left(r^4\,\mathrm{e}^{2(\gamma-\beta)} \displaystyle{U_{,r}}\right)_{,r} &=& \displaystyle{2r^2\,\left[r^2\left(\frac{\beta}{r^2}\right)_{,r\theta} - \frac{\left(\sin^2 \theta\,\gamma\right)_{,r \theta}}{\sin^2 \theta} + 2 \gamma_{,r} \gamma_{,\theta}\right]}, \label{eq5}\\
\nonumber \\
V_{,r} &=& -\frac{1}{4}r^4 \mathrm{e}^{2(\gamma-\beta)}\left(U\right)_{,r}^2 + \frac{\left(r^4 \sin \theta\,U \right)_{,r\theta}}{2 r^2 \sin \theta} \nonumber\\
    &&+ \mathrm{e}^{2(\beta-\gamma)}\bigg[1-\frac{(\sin \theta\,\beta_{,\theta})_{,\theta}}{\sin \theta} +\gamma_{,\theta\theta} + 3 \cot \theta\,\gamma_{,\theta} \nonumber \\
	\nonumber \\
	&&- (\beta_{,\theta})^2 - 2\gamma_{,\theta}(\gamma_{,\theta}-\beta_{,\theta})\bigg], \label{eq6} \\
\nonumber \\
4r(r\gamma)_{,ur} &=& \Bigg\{2r\gamma_{,r}V - r^2\left[ 2\gamma_{,\theta}U + \sin \theta\,\left(\frac{U}{\sin \theta}\right)_{,\theta}\right]\Bigg\}_{,r} \nonumber\\
&&-2 r^2\frac{(\gamma_{,r}U\sin \theta)_{,\theta}}{\sin \theta} + \frac{1}{2}r^4\mathrm{e}^{2(\gamma-\beta)} (U_{,r})^2 \nonumber \\
&&+ 2\mathrm{e}^{2(\beta-\gamma)}\bigg[(\beta_{,\theta})^2 + \sin \theta \left(\frac{\beta_{,\theta}}{\sin \theta}\right)_{,\theta}\bigg].   \label{eq7}
\end{eqnarray}
%\end{widetext}

\noindent The subscripts $u,r$ and $\theta$ denote derivatives with respect to these coordinates. In the seminal work of Bondi et al.\cite{bondi}, the characteristic formulation of General Relativity was introduced and studied for the first time. As such, the evolution scheme obeys a nice hierarchical structure: once the initial data $\gamma(u_0,r,\theta) =\gamma_0(r,\theta)$ is fixed, the hypersurface equations (\ref{eq4}) - (\ref{eq6}) determine the metric functions $\beta,U$ and $V$ (modulo integration constants) on the initial null cone $u=u_0$. Taking into account these results, the evolution equation (\ref{eq7}) provides $\gamma_{,u}$ at the initial null cone $u=u_0$, and consequently allows the determination of $\gamma$ at the next null surface $u=u_0+\delta u$. By repeating the whole cycle, we ended up with the evolution of spacetime. 

By inspecting the field equations, the conditions of regularity of the spacetime at the origin $r=0$ are 

\begin{eqnarray}
\gamma = \mathcal{O}(r^2),\; \beta = \mathcal{O}(r^4),\; U = \mathcal{O}(r),\; V = r+\mathcal{O}(r^3), \label{eq8}
\end{eqnarray}

\noindent and the conditions of smootheness on the symmetry axis ($\theta =0,\pi$) demand that

\begin{eqnarray}
\bar{\gamma} \equiv \frac{\gamma}{\sin^2 \theta}\;\; \mathrm{and}\;\; \bar{U} \equiv \frac{U}{\sin \theta}, \label{eq9}
\end{eqnarray}

\noindent are continuous functions at $\theta=0,\pi$. Taking into consideration Eq. (\ref{eq4}) and the above relation, we have 

\begin{eqnarray}
\bar{\beta} \equiv \frac{\beta}{\sin^4 \theta}. \label{eq10}
\end{eqnarray}

\noindent Then, when trackling the field equations numerically, we will consider the metric functions $\bar{\gamma}$, $\bar{U}$, $\bar{\beta}$ and $S$ defined in Ref. \cite{papadopoulos}

\begin{eqnarray}
S \equiv \frac{V-r}{r^2}.  \label{eq11}
\end{eqnarray}

\noindent In this case, the function $S$ satisfies the same condition of the metric function $\bar{U}$ with respect to the radial dependence, that is $S = \mathcal{O}(r)$.

Another relevant piece of information regards the asymptotic form of the metric functions. We demand the spacetime to be asymptotically flat, and assuming, for the sake of convenience, the Winicour-Tamburino frame \cite{winicour_tamburino,isaacson} that consists in reproducing the Minkowski metric at the center of symmetry, the following conditions must be fulfilled:

\begin{eqnarray}
\gamma &=& K(u,\theta) + \displaystyle{\frac{c(u,\theta)}{r}} + \mathcal{O}(r^{-2}), \label{eq12} \\ 
\nonumber \\
\beta &=& H(u,\theta) + \mathcal{O}(r^{-2}), \label{eq13} \\ 
\nonumber \\
U &=& L(u,\theta) + \mathcal{O}(r^{-1}),\label{eq14} \\ 
\nonumber \\
V &=& \frac{(L\sin \theta)_{,\theta}}{\sin \theta} r^2 + r \mathrm{e}^{2(H-K)} V_1(u,\theta) - 2 \mathrm{e}^{2 H} M(u,\theta) \nonumber \\
&& + \mathcal{O}(r^{-1}), \label{eq15}     
\end{eqnarray}

\noindent where $V_1(u,\theta)$ is related to the functions $H, K$ and $L$; $M(u,\theta)$ is the Bondi mass aspect \cite{bondi,papadopoulos}. It is worth mentioning that the above asymptotic expansions do not belong to the standard Bondi frame, where the functions $K,H,L$ vanish \cite{bondi}. Regarding the function $S$ its behavior is similar to the function $U$.

The Bondi formula is the main result presented in Ref. \cite{bondi}. It relates the loss of mass of a localized mass distribution due to the gravitational wave extraction. In the present frame or gauge, the Bondi formula reads as 
\begin{eqnarray}
\frac{d M_B}{d u} = -\frac{1}{2}\,\int_0^\pi\,\frac{\mathrm{e}^{2H}}{\omega}\,\mathcal{N}^2 \sin \theta\, d\theta, \label{eq16}
\end{eqnarray}

\noindent where $M_B(u)$ is the Bondi mass, $\mathcal{N}(u,\theta)$ is the news function and $\omega(u,\theta)$ is a function belonging on the gauge we are considering. We present in the Appendix the corresponding expressions for the Bondi mass and the news function.

%%%%%%%%%%%%%%%%%%%%%%%%%%%%%%%%%%%%%%%%%%%%%%%%%%%%%%%%%%%%
\section{The domain decomposition Galerkin-Collocation method}\label{Section_III}
%%%%%%%%%%%%%%%%%%%%%%%%%%%%%%%%%%%%%%%%%%%%%%%%%%%%%%%%%%%%

%\vspace{0.2cm}
%\noindent - \textit{Preliminary definitions: new functions, mapps and variables.}
%\vspace{0.10cm}

\subsection{Preliminary definitions: new variables and mapps}

Before presenting the description of the domain decomposition GC method, it will be convenient to introduce an additional function $\bar{Q}=\bar{Q}(u,r,\theta)$ by

\begin{eqnarray}
\bar{Q} = \mathrm{e}^{2(\gamma-\beta)} r^2 \bar{U}_{,r}. \label{eq17}
\end{eqnarray}

\noindent As a consequence, the second-order hypersurface equation (\ref{eq5}) is split into two first-order equations: the above equation for the metric function $\bar{U}$ and the resulting equation for the function $\bar{Q}$ after substituting Eq. (\ref{eq17}) into Eq. (\ref{eq5}). After a simple inspection, one can shown that the auxiliary function $\bar{Q}$ satisfies the same boundaries conditions of the metric function $\bar{\gamma}$. 

We divide the spatial domain $\mathcal{D}$ covered by $0 \leq r < \infty,\,0 \leq \theta \leq \pi$ into two subdomains denoted by $\mathcal{D}_1$ and $\mathcal{D}_2$:

\begin{eqnarray}
&&\mathcal{D}_1:\, 0 \leq r \leq r^{(1)},\,0 \leq \theta \leq \pi, \nonumber \\
\nonumber \\
&&\mathcal{D}_2:\, r^{(1)} \leq r \leq \infty,\,0 \leq \theta \leq \pi, \nonumber
\end{eqnarray}   

\noindent where $r=r^{(1)}$ is the interface between both subdomains. For the sake of convenience, we introduce a new variable $x$ by

\begin{equation}
x = \cos \theta, \label{eq18}
\end{equation}

\noindent meaning that the angular patch is covered by $-1 \leq x \leq 1$.

We have followed a similar strategy presented in the spherical case \cite{alcoforado_critical,alcoforado_cauchy}; namely, we first introduce an intermediary computational variable $y$ using the algebraic map \cite{boyd}

\begin{eqnarray}
r = L_0 \frac{(1+y)}{1-y}, \label{eq19}
\end{eqnarray}

\noindent where $L_0$ is the map parameter. Then, we have that $0 \leq r < \infty$ is mapped out into $-1 \leq y \leq 1$. And in second place, we introduce the following linear transformations to define the variables $-1 \leq \xi^{(l)} \leq 1$, $l=1,2$ that cover the radial sector of the subdomains:

\begin{eqnarray}
y(\xi^{(l)})=\frac{1}{2}\,\left[\left(y^{(l)}-y^{(l-1)}\right)\xi^{(l)}+y^{(l)}+y^{(l-1)}\right], \label{eq20}
\end{eqnarray} 

\noindent where $y^{(0)}=-1$ and $y^{(2)}=1$. In terms of the new variables, we have $\mathcal{D}_l:\,-1 \leq \xi^{(l)} \leq 1$ and $-1 \leq x \leq 1$ with $l=1,2$. Fig. 1 illustrates the present scheme of domain decomposition. We define the collocation points $(\xi^{(l)}_k,x_j)$ using these variables and then mapped back to $(r_k,\theta_j)$ in the physical domain $\mathcal{D}$.

It will be useful for the definition of the rational Chebyshev polynomials \cite{boyd} the relation between the radial coordinate $r$ and the computational variables $\xi^{(l)}$ for $l=1,2$. By combining Eqs. (\ref{eq19}) and (\ref{eq20}), we obtain

\begin{eqnarray}
\xi^{(l)}=\frac{a^{(l)}r+b^{(l)}}{(r+L_0)}, \label{eq21}
\end{eqnarray}

\noindent where $l=1,2$ and
\begin{eqnarray}
a^{(l)}&=&\frac{2L_0+r^{(l)}+r^{(l-1)}}{r^{(l)}-r^{(l-1)}} \label{eq22}\\
\nonumber \\ 
b^{(l)}&=&-\frac{2r^{(l)}r^{{(l-1)}}+L_0\left(r^{(l)}+r^{(l-1)}\right)}{r^{(l)}-r^{(l-1)}}. \label{eq23}
\end{eqnarray}

\noindent Here $r^{(0)}=0$ and $r^{(2)}$ is located at infinity. Notice that $r^{(l-1)}\leq r \leq r^{(l)}$ corresponds to $-1 \leq \xi^{(l)} \leq 1$ for $l=1,2$.

\begin{figure*}[htb]
\vspace{-1cm}
	\includegraphics[height=12cm,width=12cm]{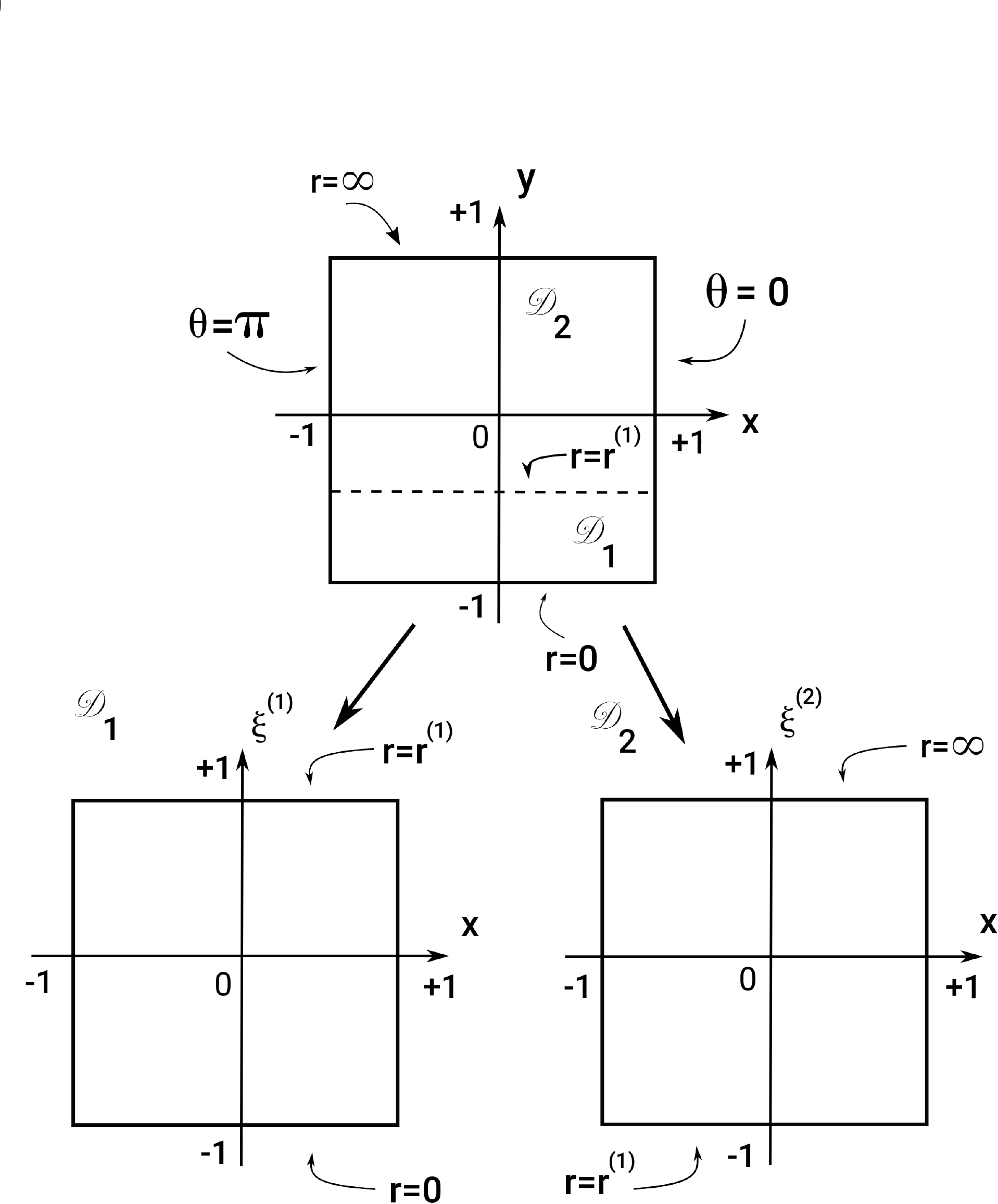}
		\caption{We illustrate in the first panel the subdomains $\mathcal{D}_1:\,0 \leq r \leq r^{(1)},\,0 \leq \theta \leq \pi$ and $\mathcal{D}_2:\,r^{(1)} \leq r < \infty,\,0 \leq \theta \leq \pi$. The radial domain $0 \leq r < \infty$ is mapped out into $-1 \leq y \leq 1$ using the map (\ref{eq19}), and the angular domain is covered by $-1 \leq x \leq 1$ with $x=\cos \theta$. In the lower panels, we have introduced the computational variables by $-1 \leq \xi_{j} \leq 1$, $j=1,2$, corresponding to the compactified radial parts of both subdomains, $-1 \leq y \leq y^{(1)}$ and $y^{(1)} \leq y \leq 1$, where $y=y^{(1)}$, $-1 \leq x \leq 1$ represents the interface.}
\end{figure*}

\subsection{Spectral approximations}
We establish the following spectral approximations of the metric functions  $\bar{\gamma}(u,r,\theta),\,\bar{\beta}(u,r,\theta),\,\bar{U}(u,r,\theta),\,\bar{Q}(u,r,\theta)$ and $S(u,r,\theta)$ in each subdomain:
\begin{eqnarray}
\bar{\gamma}^{(l)}(u,r,\theta)&=&\sum^{N_\gamma^{(l)}}_{k=0}\sum^{\tilde{N}_\gamma^{(l)}}_{j=0}\,a^{(l)}_{kj}(u)\psi^{(l)}_k(r)P_j(\cos \theta), \label{eq24}\\
\nonumber \\
\bar{\beta}^{(l)}(u,r,\theta)&=&\sum^{N_\beta^{(l)}}_{k=0}\sum^{\tilde{N}_\beta^{(l)}}_{j=0}\,c^{(l)}_{kj}(u)\Phi^{(l)}_k(r)P_j(\cos \theta), \label{eq25}\\
\nonumber \\
\bar{U}^{(l)}(u,r,\theta)&=&\sum^{N_U^{(l)}}_{k=0}\sum^{\tilde{N}_U^{(l)}}_{j=0}\,b^{(l)}_{kj}(u)\chi^{(l)}_k(r)P_j(\cos \theta), \label{eq26}\\
\nonumber \\
\bar{Q}^{(l)}(u,r,\theta)&=&\sum^{N_Q^{(l)}}_{k=0}\sum^{\tilde{N}_Q^{(l)}}_{j=0}\,q^{(l)}_{kj}(u)\psi^{(l)}_k(r)P_j(\cos \theta), \label{eq27}\\
\nonumber \\
S^{(l)}(u,r,\theta)&=&\sum^{N_S^{(l)}}_{k=0}\sum^{\tilde{N}_S^{(l)}}_{j=0}\,s^{(l)}_{kj}(u)\chi^{(l)}_k(r)P_j(\cos \theta), \label{eq28}
\end{eqnarray}

\noindent where $l=1,2$ indicates the specific subdomain; $\{a^{(l)}_{kj}(u),\,b^{(l)}_{kj}(u),\,c^{(l)}_{kj}(u),\,q^{(l)}_{kj}(u),\,s^{(l)}_{kj}(u)\}$ are the modes or the unknown coefficients. The number of modes in each subdomain depends on the radial and angular truncation orders denoted generically by $N^{(l)}$ and $\tilde{N}^{(l)}$, respectively. The angular basis functions common in each subdomain are the Legendre polynomials, $P_j(\cos \theta)$, whereas the radial basis functions $\{\psi^{(l)}_k(r),\,\Phi^{(l)}_k(r),\,\chi^{(l)}_k(r)\}$ are constructed to satisfy the conditions at $r=0$ and $r=\infty$. This is possible if we combine appropriately the rational Chebyshev polynomials defined in each subdomain as shown in the sequence.

\subsection{Radial basis functions}

The rational Chebyshev polynomials are defined in each subdomains as
\begin{eqnarray}
TL_k^{(l)} = T_k\left(\xi^{(l)}=\frac{a^{(l)}r+b^{(l)}}{(r+L_0)}\right), \label{eq29}
\end{eqnarray}

\noindent where $T_k(\xi)$ is the Chebyshev polynomial of kth order; the parameters $a^{(l)}$ and $b^{(l)}$ are given by the relations (\ref{eq22}) and (\ref{eq23}), respectively.

We can now define the radial basis functions for the spectral approximations (\ref{eq24}) - (\ref{eq28}). As we have mentioned, the radial basis functions must satisfy the conditions near the origin and infinity ($\mathcal{J}^+$) provided by the relations (\ref{eq8}) and (\ref{eq12}) - (\ref{eq15}). 

We start with the common radial basis for the functions $\bar{U}$ and $S$ in the first subdomain $\mathcal{D}_1$. Since near the origin $\bar{U}=\mathcal{O}(r)$, it follows that
\begin{eqnarray} 
\chi_k^{(1)} = \frac{1}{2}\,\left(TL_{k+1}^{(1)}(r)+TL_{k}^{(1)}(r)\right). \label{eq30}
\end{eqnarray} 

\noindent The radial basis functions for $\bar{\gamma}$ and $\bar{Q}$ in the first subdomains is a simple linear combination of the aboves basis functions:
\begin{eqnarray} 
\psi_k^{(1)} = \frac{1}{4}\,\left[\frac{(2k+1)}{(2k+3)}\,\chi_{k+1}^{(1)}(r)+\chi_{k}^{(1)}(r)\right], \label{eq31}
\end{eqnarray} 

\noindent that satisfy $\psi_k^{(1)}(r) = \mathcal{O}(r^2)$ near the origin for all $k$. The radial basis function for $\bar{\beta}$ is an elaborated combination of the above basis. The basis functions $\Phi_k^{(1)}(r)$ are defined by 
\begin{eqnarray} 
\Phi_k^{(1)} &&= \frac{1}{2}\,\frac{(k+1)(2k+1)(2k+3)}{(2k+5)^2(k+3)}\,\psi_{k+2}^{(1)}(r)\nonumber \\
\nonumber \\
&&+2\frac{(k+1)(2k+1)}{(2k+5(2k+3))}\psi_{k+1}^{(1)}(r)+\frac{1}{2}\psi_k^{(1)}(r).  \label{eq32}
\end{eqnarray} 

\noindent In this case, we have $\Phi_k^{(1)}(r) = \mathcal{O}(r^4)$ for all $k$ as required by (\ref{eq8}). 

In the second subdomain, we choose the rational Chebyshev polynomials defined in $\mathcal{D}_2$ as the radial basis functions for all fields. We summarize these results in Table 1.

\begin{table}
	\centerline{Table 1}
	\medskip
	\begin{center}
		\centering
		\begin{tabular}[c]{ l|c|c }
			\hline
			%\\
			&$\mathcal{D}_1$ & $\mathcal{D}_2$  \\
			%\\
			\hline
			\hline
			%& &\\
			$\bar{\gamma},\,\bar{Q}$ & $\phi_k^{(1)}(r)$  & $  $ \\
			& & \\
			%\vspace{0.2cm}
			%\hline
			$\bar{\beta}$ & $\Phi_k^{(1)}(r)$ & $TL_k^{(2)}(r)$\\
			& & \\
			%\hline
			%\\
			$\bar{U},\,S$ & $\chi_k^{(1)}(r)$&   \\
			\hline
		\end{tabular}
		\caption{Radial basis functions for the spectral approximations (\ref{eq24}) - (\ref{eq28}).}
	\end{center}
\end{table}

\subsection{Transmission conditions}

The next step is to establish the transmission conditions to guarantee that all pieces of the spectral approximations (\ref{eq24}) - (\ref{eq28}) represent the same corresponding functions, i. e. $\bar{\gamma}, \bar{\beta}, \bar{U}, \bar{Q}$ and $S$ defined in each subdomain. As in the spherical case \cite{alcoforado_critical,alcoforado_cauchy}, we adopt the patching method \cite{canuto} that demands that a function and all their $d-1$ spatial derivatives must be continuous at the contiguous subdomains' interface. Since the interface between both subdomains is defined by $r-r^{(1)} = 0,\,0 \leq \theta \leq \pi$, the spatial derivatives involves only the radial coordinate $r$. After inspecting the field equations (\ref{eq4}) - (\ref{eq7}), we come up with the following transmission conditions: 

\begin{eqnarray}
\bar{\beta}^{(1)}(u,r^{(1)},\theta)&=&\bar{\beta}^{(2)}(u,r^{(1)},\theta), \label{eq33}\\
\nonumber \\
\bar{U}^{(1)}(u,r^{(1)},\theta)&=&\bar{U}^{(2)}(u,r^{(1)},\theta),\label{eq34}\\
\nonumber \\
\bar{Q}^{(1)}(u,r^{(1)},\theta)&=&\bar{Q}^{(2)}(u,r^{(1)},\theta), \label{eq35}\\
\nonumber \\
S^{(1)}(r,r^{(1)},\theta)&=&S^{(2)}(r,r^{(1)},\theta), \label{eq36}\\
\nonumber \\ \bar{\gamma}^{(1)}(u,r^{(1)},\theta)&=&\bar{\gamma}^{(2)}(u,r^{(1)},\theta), \label{eq37}\\
\nonumber \\
\left(\frac{\partial \bar{\gamma}^{(1)}}{\partial r}\right)_{r^{(1)}}&=&\left(\frac{\partial \bar{\gamma}^{(2)}}{\partial r}\right)_{r^{(1)}}, \label{eq38}
\end{eqnarray}

\noindent that are valid for $0 \leq \theta \leq \pi$. In the numerical implementation, the above conditions are to be taken approximately.

\subsection{Implementing the GC domaind decomposition method}
	
The field equations under consideration are formed by four hypersurface equations for the functions $\bar{\beta},\bar{Q},\bar{U}$ and $S$ and an wave equation for $\bar{\gamma}$. In this case, any spectral method will approximate the hypersurface equations into sets of algebraic equations for the corresponding modes and a set of ordinary differential equations for the modes $a_{kj}^{(l)}(u)$,  $k=0,1..,N_\gamma^{(l)}, j=0,1..,\tilde{N}_\gamma^{(l)}$ with $l=1,2$. In the sequence, we summarize the description of the procedure that follows closely our previous references \cite{alcoforado_critical,alcoforado_cauchy,barreto_DD, crespo_affine}.

\subsubsection{Hypersurface equations for $\beta$, $Q$ and $U$.}

Let us consider the hypersurface equation for $\bar{\beta}$, Eq. (\ref{eq4}). By substituting the spectral approximations for $\bar{\gamma}$ and $\bar{\beta}$ (Eqs. (\ref{eq24}) and (\ref{eq25}), respectively), we obtain the corresponding residual equation

\begin{eqnarray}
%\mathrm{Res}_\beta^{(l)}(u,r,\theta) = %\sum_{m,n=0}^{M^{(l)}_\beta,N^{(l)}_\beta}c^{(l)}_{mn} %\Phi_{m,r}^{(l)}(r)P(\cos \theta) - 
%\frac{1}{2}r(\bar{\gamma}_{,r}^{(l)})^2, \nonumber \\
\mathrm{Res}_\beta^{(l)}(u,r,\theta) = \bar{\beta}_{,r}^{(l)} - \displaystyle{\frac{1}{2} r\left(\bar{\gamma}_{,r}^{(l)}\right)^2},\label{eq39}
\end{eqnarray}

\noindent where $l=1,2$. These equations do not vanish due to the introduced spectral approximations. Next, to obtain a set of equations for the modes $c_{mn}^{(l)}(u)$, $m=0,1,..,N_\beta^{(l)}, n=0,1,..,\tilde{N}_\beta^{(l)}$ and $l=1,2$, we follow the prescription of the Collocation method that consists in vanishing the residual equations (\ref{eq39}) at the collocation points $(r^{(l)}_k,\theta_j)$ in each subdomain. Equivalently, the test functions are deltas of Dirac, $\delta(r-r^{(l)}_k) \delta(\theta-\theta_j)$. Then, we have   

\begin{eqnarray}
(\bar{\beta}_{,r}^{(l)})_{kj} -  \frac{1}{2}r^{(l)}_k\,(\bar{\gamma}_{,r}^{(l)})_{kj}^2 = 0,\;l=1,2. \label{eq40}
\end{eqnarray}

\noindent The quantities where $(\bar{\beta}_{,r}^{(l)})_{kj}$ and $(\bar{\gamma}_{,r}^{(l)})_{kj}$ are the values of the derivatives of $\bar{\beta}^{(l)}$ and $\bar{\gamma}^{(l)}$ at the collocation points:
\begin{eqnarray}
(\bar{\beta}_{,r}^{(l)})_{kj} = \sum_{m,n}\,c_{mn}^{(l)}(u)\left(\frac{\partial \Phi_m^{(l)}}{\partial r}\right)_{r^{(l)}_k}\,P_n(\cos \theta_j),\label{eq41} \\
\nonumber \\
(\bar{\gamma}_{,r}^{(l)})_{kj} = \sum_{m,n}\,a_{mn}^{(l)}(u)\left(\frac{\partial \psi_m^{(l)}}{\partial r}\right)_{r^{(l)}_k}\,P_n(\cos \theta_j).\label{eq42}
\end{eqnarray}

\noindent We have to establish the collocation points in each subdomain such that their number together with the transmission conditions must be equal to 

\[(N_\beta^{(1)}+1)(\tilde{N}_\beta^{(1)}+1)+(N_\beta^{(2)}+1)(\tilde{N}_\beta^{(2)}+1),\]

\noindent that is the total number of modes $c_{mn}^{(l)}$.

To obtain the approximate equations from the transmission condition (\ref{eq33}), we first need to write the corresponding residual equation
\begin{eqnarray}
\mathrm{Res}_\beta^{(TC)}(u,\theta) = \bar{\beta}^{(1)}(u,r^{(1)},\theta) - \bar{\beta}^{(2)}(u,r^{(1)},\theta).  \label{eq43}
\end{eqnarray} 

\noindent We can make the above equation to vanish at the angular collocation points $\theta=\theta_j$, or alternatively to impose that $\mathrm{Res}_\beta^{(TC)}(u,\theta)$ vanish in an average sense, meaning that

\begin{eqnarray}
\left<\mathrm{Res}_\beta^{(TC)}(u,\theta),P_j(\cos \theta)\right> = \int_{-1}^1\,\mathrm{Res}_\beta^{(TC)}(u,x) P_j(x) dx = 0, \nonumber \\ \label{eq44}
\end{eqnarray}
	
\noindent for all $j=0,1,..,\tilde{N}_\beta^{(2)}$. Therefore, we have $\tilde{N}_\beta^{(2)}+1$ equations related to the transmission condition (\ref{eq33}) demanding 

\[(N_\beta^{(1)}+1)(\tilde{N}_\beta^{(1)}+1)+N_\beta^{(2)}(\tilde{N}_\beta^{(2)}+1),\]

\noindent collocation points $(r_k^{(l)},x_j),\,l=1,2$. Based on the above considerations, we choose the following collocation points in the first subdomain:

\begin{eqnarray}
\mathcal{D}_1 : \begin{cases}
\xi^{(1)}_k = \displaystyle{\cos\left(\frac{k \pi}{N_\beta^{(1)}+2}\right)},\;k=1,2,..,N_\beta^{(1)}+1,\\
\\
x_j = -1,\,\mbox{zeros of}\,\,\displaystyle{\frac{d P_{\tilde{N}_\beta^{(1)}}}{dx}},1,\;j=0,1,..,\tilde{N}_\beta^{(1)},
\end{cases} \label{eq45}
\end{eqnarray}

\noindent where $\xi_0^{(1)}=1$ and $\xi_{N_\beta^{(1)}+2}^{(1)}=-1$ correspond to the interface $r=r^{(1)}$ and the origin, $r=0$, respectively, that are not included in the first subdomain. For the second subdomain, we have 

\begin{eqnarray}
\mathcal{D}_2: \begin{cases} 
\xi^{(2)}_k = \displaystyle{\cos\left(\frac{k \pi}{N_\beta^{(2)}}\right)},\;k=1,2,..,N_\beta^{(2)},\\
\\
x_j = -1,\,\mbox{zeros of}\,\,\displaystyle{\frac{d P_{\tilde{N}_\beta^{(2)}}}{dx}},1,\;j=0,1,..,\tilde{N}_\beta^{(2)}.
\end{cases} \label{eq46}
\end{eqnarray}

\noindent Here $\xi_0^{(2)}=1$ corresponds to the infinity $\mathcal{J}^+$ and is excluded since the residual equation vanish identically; $\xi^{(2)}_{N_\beta^{(2)}}=-1$ is the interface $r=r^{(1)}$.

Now, we can summarize the determination or the update to the modes $c_{mn}^{(l)}$ from the residual equations (\ref{eq40}) together with the transmission conditions (\ref{eq44}). It is convenient to substitute the two indexes that characterize the values $(\bar{\beta}^{(l)}_{,r})_{kj}$ as well as the modes $c^{(l)}_{mn}$ by one index, or 

\[(\bar{\beta}^{(l)}_{,r})_{kj} \, \rightarrow \, (\bar{\beta}_{,r})_i,\quad c^{(l)}_{mn} \, \rightarrow \, c_i.\]

\noindent For instance, $(\bar{\beta}^{(1)}_{,r})_{11}=(\bar{\beta}_{,r})_1,\,(\bar{\beta}^{(1)}_{,r})_{12}=(\bar{\beta}_{,r})_2,...$ and so on. The relation between the values $(\bar{\beta}_{,r})_i$ and the modes $c_i$ is written in a matrix form

\begin{eqnarray}
\begin{pmatrix}
	(\bar{\beta}_{,r})_1\\
	(\bar{\beta}_{,r})_2\\
	\vdots\\
	(\bar{\beta}_{,r})_{K_\beta}\\
	0\\
	\vdots \\
	0	
\end{pmatrix}
=\mathbb{B}\,
\begin{pmatrix}
c_1\\
c_2\\
\vdots\\
c_{K_\beta}\\
c_{K_\beta+1}\\
\vdots \\
c_{N_\beta}	
\end{pmatrix}, \label{eq47}
\end{eqnarray}

\noindent where the elements of the matrix $\mathbb{B}_{pq}$ are given by

\[\left(\frac{d\Phi_m^{(l)}}{d r}\right)_{r_k}P_n(x_j), \]

\noindent with the convention $kj \rightarrow p$ and $mn \rightarrow q$ (including $l=1,2$). Still,  $K_\beta=(N_\beta^{(1)}+1)(\tilde{N}_\beta^{(1)}+1)+N_\beta^{(2)}(\tilde{N}_\beta^{(2)}+1)$ is the number of collocation points and $N_\beta=K_\beta+\tilde{N}_\beta^{(2)}+1$ is the number of modes $c_{mn}^{(l)},\,l=1,2$. In the above matricial equation, the zeros on the first column vector correspond the inclusion of the transmission conditions. Thus, the determination of modes $c_i$ is obtained after inverting the matrix $\mathbb{B}$ with the values $(\bar{\beta}_{,r})_j$ determined from the relations (\ref{eq40}).% since the values $(\bar{\gamma}_{,r}^{(l)}_j$.

We have implemented a similar procedure with respect to the equations for $\bar{Q}$ and $\bar{U}$. The only difference is the number of matrices relating the values of $\bar{\beta}_{kj},(\bar{\beta}_{,x})_{kj},(\bar{\beta}_{,xr})_{kj},\bar{\gamma}_{kj},(\bar{\gamma}_{,r})_{kj},(\bar{\gamma}_{,xr})_{kj}$ and $(\bar{\gamma}_{,xr})_{kj}$ (we have dropped the index that indicates the subdomain for simplicity). These matrices and the hypersurface equations for $\bar{Q}$ and $\bar{U}$ determine the values $(\bar{Q}_{,r})_{kj}$ and $(\bar{U}_{,r})_{kj}$ that together with the corresponding transmission conditions, we can update the modes $q_{kj},\,b_{kj}$ in the same way as we have implemented previously for the modes $c_{kj}$. For the sake of simplicity, have considered the same collocation points with coordinates (\ref{eq45}) and (\ref{eq46}), in other words we are assuming that $N_\beta^{(l)}=N_Q^{(l)}=N_U^{(l)}$ and $\tilde{N}_\beta^{(l)}=\tilde{N}_Q^{(l)}=\tilde{N}_U^{(l)}$, $l=1,2$.

%The corresponding hypersurfaces equations evaluated at the collocation points

\subsubsection{Hypersurface equation for $S$ and the evolution equation}

In the sequence, we have considered the hypersurface equation for $S$ and the evolution equation, Eqs. (\ref{eq6}) and (\ref{eq7}), respectively. The common aspect shared in treating both equations spectrally is a new set of collocation points determined by the radial and angular truncation orders of $S$, $N_S^{(l)}$ and $\tilde{N}_S^{(l)}$, respectively, where, in general, we have $N_S^{(l)} \geq N_\beta^{(l)}$ and $\tilde{N}_S^{(l)} \geq \tilde{N}_\beta^{(l)}$, $l=1,2$.

The new set of grid points defined in the first and second subdomains is

\begin{eqnarray}
\mathcal{D}_1 : \begin{cases}
\xi^{(1)}_k = \displaystyle{\cos\left(\frac{k \pi}{N_S^{(1)}+2}\right)},\;k=1,2,..,N_S^{(1)}+1,\\
\\
x_j = -1,\,\mbox{zeros of}\,\,\displaystyle{\frac{d P_{\tilde{N}_S^{(1)}}}{dx}},1,\;j=0,1,..,\tilde{N}_S^{(1)}.
\end{cases} \label{eq48}
\end{eqnarray}
%\noindent And in the second subdomain, we have
%
\begin{eqnarray}
\mathcal{D}_2 : \begin{cases}
\xi^{(2)}_k = \displaystyle{\cos\left(\frac{k \pi}{N_S^{(2)}-1}\right)},\;k=0,1,..,N_S^{(2)}-1,\\
\\
%\nonumber \\
x_j = -1,\,\mbox{zeros of}\,\,\displaystyle{\frac{d P_{\tilde{N}_S^{(2)}}}{dx}},1,\;j=0,1,..,\tilde{N}_S^{(2)}.
\end{cases} \label{eq49}
\end{eqnarray}

\noindent As in the first grid, the radial points of $\mathcal{D}_1$ do not include the origin $r=0$ and the interface $r=r^{(1)}$. On the other hand, $\xi_0^{(2)}=1$ corresponds to $r=\infty$ and $\xi_{N_S^{(2)}-1}^{(2)}=-1$ corresponds to the interface $r=r^{(1)}$, both radial points belong to the second subdomain. Therefore, we have a total of

\[(N_S^{(1)}+1)(\tilde{N}_S^{(1)}+1)+N_S^{(2)}(\tilde{N}_S^{(2)}+1)\]

\noindent collocation points and $\tilde{N}_S^{(2)}+1$ relations arising from the transmission conditions for the function $S$:

\begin{eqnarray}
\left<\mathrm{Res}_S^{(TC)}(u,\theta),P_j(\cos \theta)\right> = \int_{-1}^1\,\mathrm{Res}_S^{(TC)}(u,x) P_j(x) dx = 0, \nonumber \\ \label{eq50}
\end{eqnarray}

\noindent where $j=0,1,..,\tilde{N}_S^{(2)}$ and $\mathrm{Res}_S^{(TC)}(u,\theta) = S^{(1)}(u,r^{(1)},\theta) - S^{(2)}(u,r^{(1)},\theta)$.

To update the modes $s_{mn}^{(l)}$, $l=1,2$, we proceed in the same way  as outlined for the modes $c_{mn}^{(l)}$.  We first construct the matrices relating the values of $\bar{\gamma}_{kj},(\bar{\gamma}_{,x})_{kj},(\bar{\gamma}_{,xx})_{kj},\bar{\beta}_{kj},(\bar{\beta}_{,x})_{kj},(\bar{\beta}_{,xx})_{kj},\bar{U}_{kj},(\bar{U}_{,r})_{kj}$ and $(\bar{U}_{,xr})_{kj}$ at the second set of collocation points. These values allow the determination of the values $(S_{,r})_{kj}$ at the same collocation points from the residual equation associated to Eq. (\ref{eq6}). Together with the transmission conditions (\ref{eq50}), we can write 

\begin{eqnarray}
\begin{pmatrix}
(S_{,r})_1\\
(S_{,r})_2\\
\vdots\\
(S_{,r})_{K_S}\\
0\\
\vdots \\
0	
\end{pmatrix}
=\mathbb{S}\,
\begin{pmatrix}
s_1\\
s_2\\
\vdots\\
s_{K_S}\\
s_{K_S+1}\\
\vdots \\
s_{N_S}	
\end{pmatrix},\label{eq51}
\end{eqnarray}
%\vspace{0.1cm}

\noindent where $K_S=(N_S^{(1)}+1)(\tilde{N}_S^{(1)}+1)+N_S^{(2)}(\tilde{N}_S^{(2)}+1)$ and $N_S=K_S+\tilde{N}_S^{(2)}+1$ is the total number of modes $s_{mn}^{(l)}$ or $s_i$ after transforming two indexes into one. The matrix components $\mathbb{S}_{pq}$ are determined after following similar convention we have presented in Eq. (\ref{eq47}). 

The last step is to consider the evolution equation (\ref{eq7}). We established the corresponding residual equation, $\mathrm{Res}_\gamma^{(l)}(u,r,x)$, at each subdomain after substituting the spectral approximations (\ref{eq24}) - (\ref{eq28}) into Eq. (\ref{eq7}) yielding

%\begin{widetext}
\begin{eqnarray}
\mathrm{Res}_\gamma^{(l)}(u,r,x) &=&(r\bar{\gamma}_{,u})_{,r} \nonumber \\
&&- \frac{1}{4r}\,\bigg[2r\bar{\gamma}_{,r}(r+r^2S) -2 (1-x^2) r^2 \bar{\gamma}_{,x} + r^2\,(4 x \bar{\gamma}\bar{U}+\bar{U}_{,x})\bigg]_{,r} \nonumber \\
&&+2xr\bar{\gamma}_{,r}\bar{U} - \frac{1}{2}(1-x^2)r(\bar{\gamma}_{,r}\bar{U})_{,x}-\frac{r^3}{8}\mathrm{e}^{2(\gamma-\beta)}\bar{U}_{,r}^2 
\nonumber \\
&&- \frac{1}{2r}\mathrm{e}^{2(\beta-\gamma)}\bigg\{(1-x^2)^2\big[4x\bar{\beta} - (1-x^2)\bar{\beta}_{,x}\big]^2\nonumber \\
&&+\left[(1-x^2)^2\bar{\beta}_{,x} - 4x(1-x^2)\bar{\beta}\right]_{,x}\bigg\}, \label{eq52}
\end{eqnarray}%\end{widetext}

\noindent where we have divided Eq. (\ref{eq7}) by $4 r \sin^2\theta$ and introduced explicitely the variable $x$. For simplicity, we have dropped the index that indicates the subdomain on the RHS of the above equation.

At this point, we can follow the procedure adopted for all residual equations, i. e. imposing the vanishing of the residual equation (\ref{eq52}) at a convenient set of collocation points. Instead, we propose to extend the strategy of Ref. \cite{rodrigues} in the case of a single domain. The procedure consists in to choose the modes $a_{mn,u}^{(l)}$ such that the residual equation (\ref{eq52}) is forced to vanish in an average sense according to the Residual Weighted Methods \cite{finlayson}. It means that the test functions are no longer Dirac functions but the same as the basis functions for $\bar{\gamma}^{(l)}$ (cf. Eq. (\ref{eq2})). Then, we evaluate the inner products using quadrature formulae as indicated by the G-NI method taking advantage of the second set of collocation points (48) and (49) as
\begin{eqnarray}
&&\displaystyle{\left<\mathrm{Res}_\gamma(u,r(\xi),x),\psi_m(y)P_n(x)\right>^{(l)}} \nonumber \\ 
\nonumber \\
&&\approx \sum_{k,j}^{N_S^{(l)},\tilde{N}_S^{(l)}}\,(\mathrm{Res}_\gamma)^{(l)}_{kj}\psi_m^{(l)}(\xi_k)P_n(x_j) w_k^{(l)} v_j^{(l)}, \label{eq53}
\end{eqnarray}

\noindent where $l=1,2$, $m=0,1,..,N_\gamma^{(l)}-1$, $n=0,1,..,\tilde{N}_\gamma^{(l)}$, and $w_k^{(l)},v_j^{(l)}$ are the weights associated to the radial and angular test functions, respectively. It is worth mentioning that we have $N_\gamma^{(1)}(\tilde{N}_\gamma^{(1)}+1)+N_\gamma^{(2)}(\tilde{N}_\gamma^{(2)}+1)$ equations relating the modes $a_{mn,u}^{(l)}$ with the values of the several quantities present in the residual evolution equation. The remaining $(\tilde{N}_\gamma^{(1)}+1)+(\tilde{N}_\gamma^{(2)}+1)$ relations are from the transmission conditions:

\begin{eqnarray}
\int_{-1}^1\,\left(\bar{\gamma}_{,u}^{(1)}(u,r^{(1)},x) - \bar{\gamma}_{,u}^{(2)}(u,r^{(1)},x)\right)\,P_j(x) dx = 0, \nonumber \\ \label{eq54}
\end{eqnarray}

\noindent where $j=0,1,..,\tilde{N}_\gamma^{(2)}$ and
\begin{eqnarray}
\int_{-1}^1\,\left(\bar{\gamma}_{,ur}^{(1)}(u,r^{(1)},x) - \bar{\gamma}_{,ur}^{(2)}(u,r^{(1)},x)\right)\,P_j(x) dx = 0, \nonumber \\ \label{eq55}
\end{eqnarray}

\noindent where $j=0,1,..,\tilde{N}_\gamma^{(1)}$. The set of equations (53), (54) and (55) can be solved to obtain the set of dynamical equations for the modes $a_{mn}^{(l)}(u)\,\rightarrow\,a_k(u)$. Schemactically, we have
\begin{eqnarray}
\frac{d a_k}{d u} = F_k\left(\bar{\gamma}_{ij},(\bar{\gamma}_{,x})_{ij},(\bar{\gamma}_{,xx})_{ij},..\right),
\end{eqnarray}

\noindent where $k=1,2,..,(N_\gamma^{(1)}+1)(\tilde{N}_\gamma^{(1)}+1)+(N_\gamma^{(2)}+1)(\tilde{N}_\gamma^{(2)}+1)$, and $\bar{\gamma}_{ij},(\bar{\gamma}_{,x})_{ij},(\bar{\gamma}_{,xx})_{ij},..$ are the values of these quantities calculated in the second grid (\ref{eq48}) and (\ref{eq49}).

\section{Code validation}

In all numerical experiments, we take advantage of the field equation's symmetry concerning the angular dependence of the functions $\bar{\gamma},\bar{\beta},\bar{U}$ and $S$. Let us consider that $\bar{\gamma}(u,r,x)$ is an even function of $x$. Eq. (\ref{eq4}) always demands that $\bar{\beta}(u,r,x)$ is an even function of $x$. After inspecting Eq. (\ref{eq5}), it follows that $\bar{U}(u,r,x)$, and consequently $\bar{Q}(u,r,x)$, must be an odd function of $x$. Inserting these informations into Eq. (\ref{eq6}), one can show that $S(u,r,x)$ is also an even function of $x$. We can define parity in $x$ in the spectral approximations (\ref{eq24}) - (\ref{eq28}) by setting the angular basis functions to $P_{2j}(x)$ and $P_{2j+1}(x)$, respectively, for the even and odd functions of $x$. A direct consequence, we increase the angular resolution by placing the collocation points $x_j$ in the interval $0 \leq x \leq 1$. %In general, $C_{\mathrm{max}$ occurs when the Bondi mass varies more rapidly, therefore being dependend on the initial amplitude. For $A_0=30, 50$ (Fig. 2) we found $C_{\mathrm{max} \approx 0.5$.

\begin{figure}[htb]
	\includegraphics[width=7.5cm,height=6.0cm]{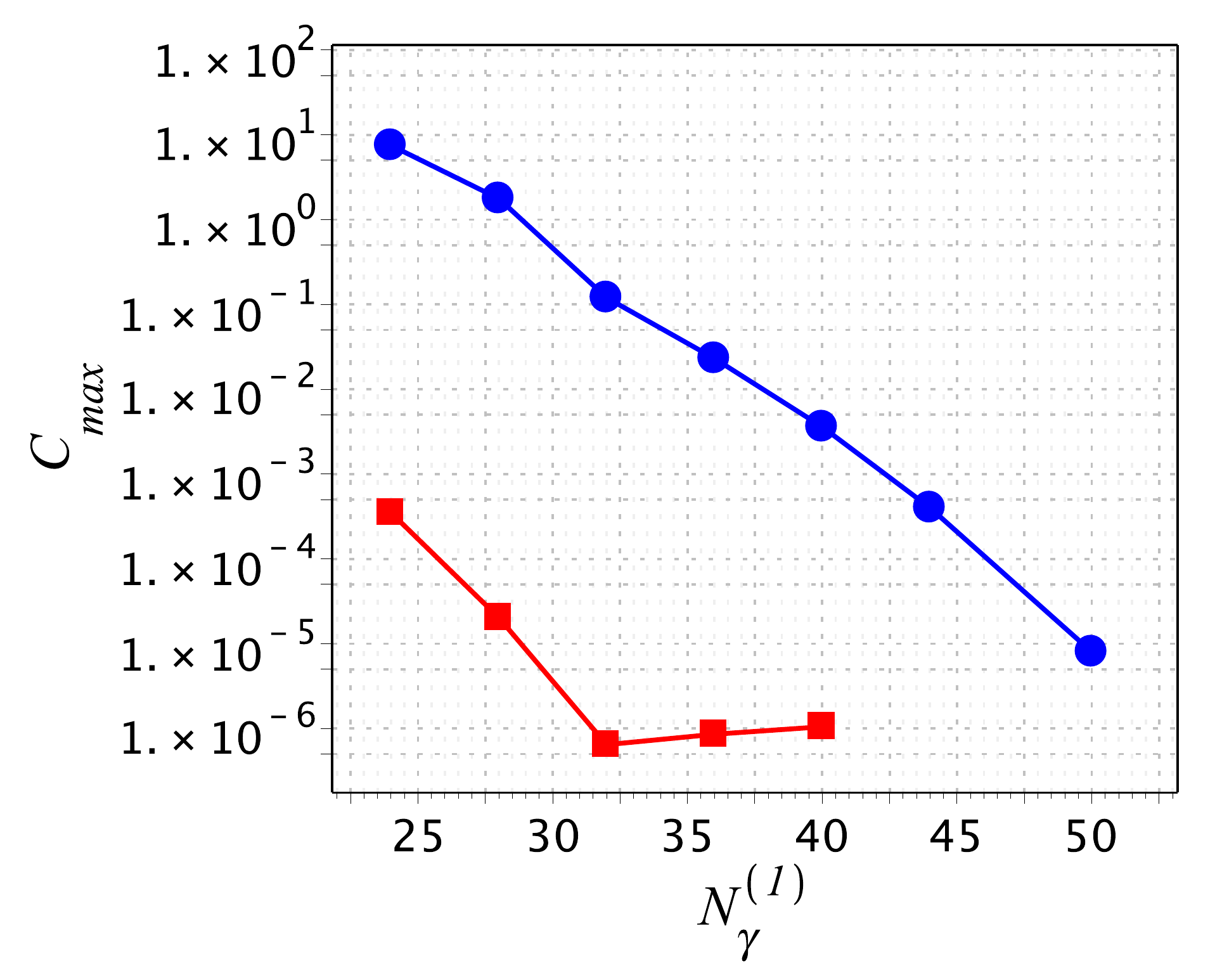}
	\caption{Exponential decay of maximum deviation $C_{\mathrm{max}}$ for the initial data (\ref{eq58}) with $A_0=30$ (squares) and $50$ (circles). $N_\gamma^{(1)}$ is the truncation order in the first domain taken as a reference. Here $y^{(1)}=0$ (cf. Eq. (\ref{eq19})) and $L_0=0.5$ for $A_0=30$, and $L_0=0.25$ for $A_0=50$. The interface is located at $r^{(1)}=0.5$ and $r^{(1)}=0.25$, respectively.}
\end{figure}

\begin{figure}[htb]
	\includegraphics[width=7.cm,height=6.0cm]{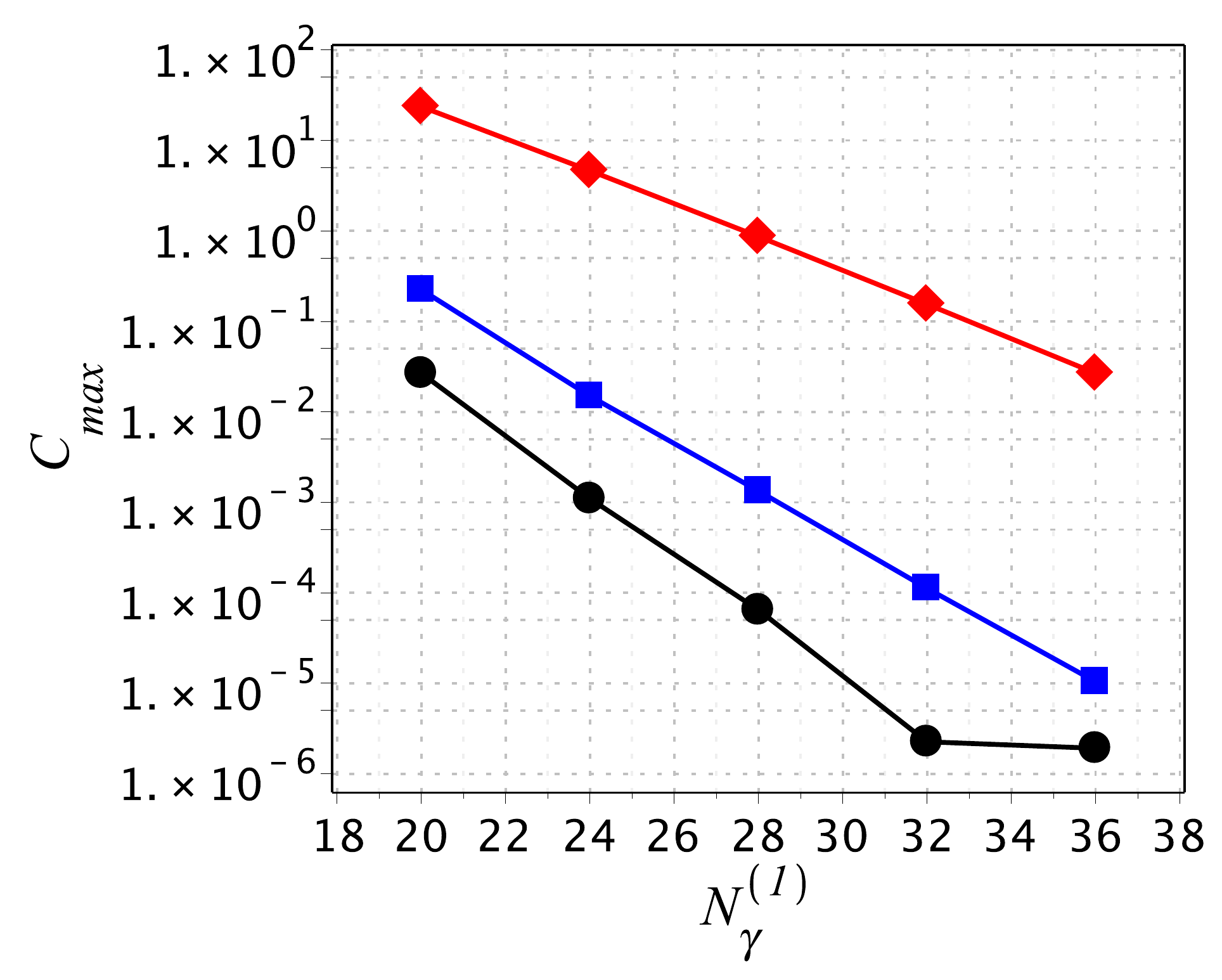}
	\includegraphics[width=7.cm,height=6.0cm]{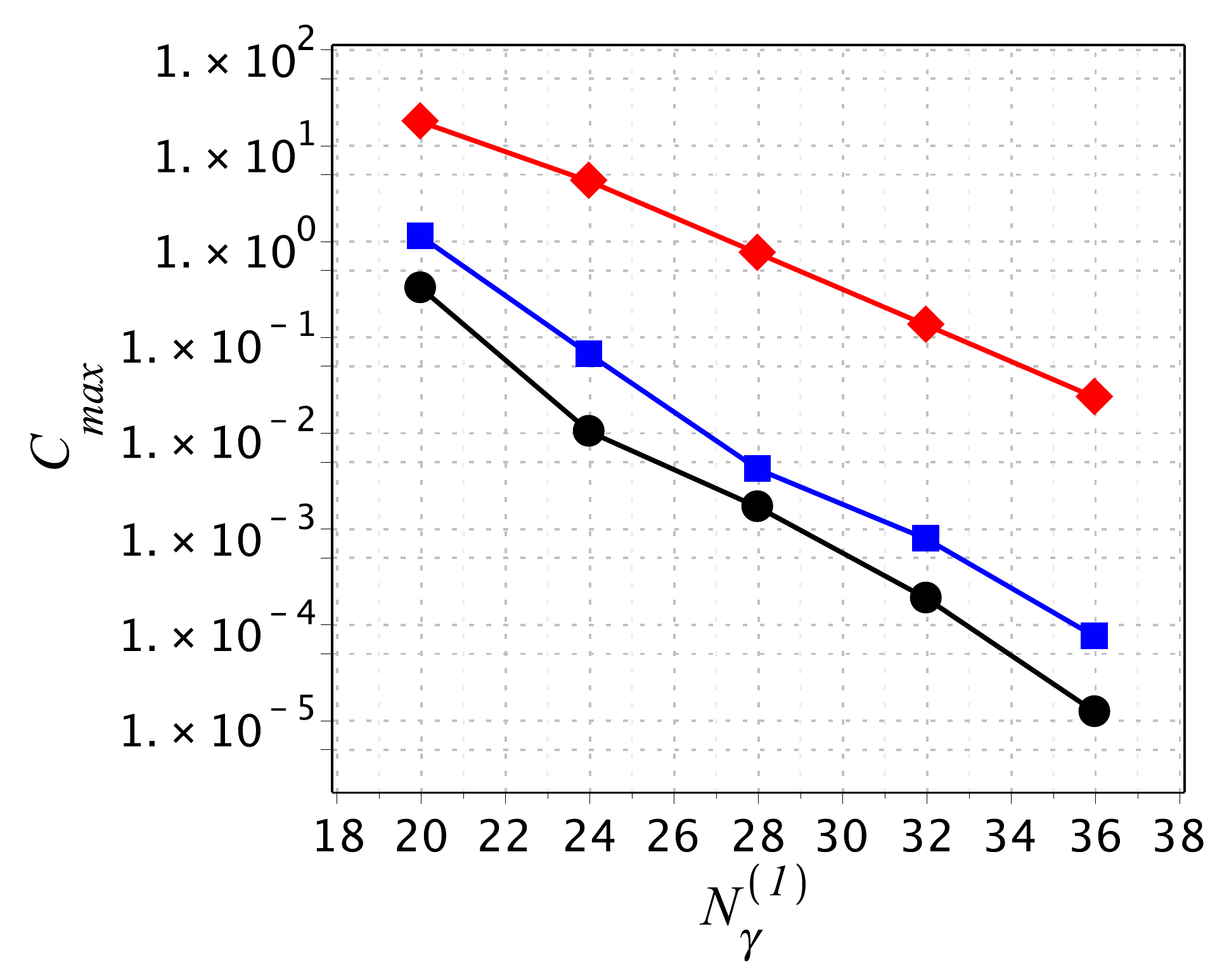}
	\caption{Exponential decay of maximum deviation $C_{\mathrm{max}}$ for the initial data (\ref{eq58}) with $A_0=30, 40$ and $50$ represented by black circles, blue squares, and red diamonds, respectively . $N_\gamma^{(1)}$ is the truncation order in the first domain taken as a reference. Here $L_0=1.0$ (upper panel) and $L_0=0.5$ (lower panel). We have set $y^{(1)}=-0.5$ (cf. Eq. (19)) yielding the interface located at $r^{(1)}=L_0/3$.}
\end{figure}

We have additionally validated the domain decomposition code with the verification of the Bondi formula. Following Gomez et al. \cite{papadopoulos}, we introduce the function $C(u)$ that measures the deviation from the Bondi formula or the global energy conservation: 

\begin{eqnarray}
C(u) = \frac{1}{M_B(u_0)}\Bigg[M_B(u)-M_B(u_0) + \frac{1}{2}\times \nonumber \\
\nonumber \\
+ \int_{u_0}^u \int_{-1}^{1}\left(\frac{\mathrm{e}^{2H}}{\omega} \mathcal{N}(u,x)^2dx\right)\,du \Bigg] \times 100, \label{eq57}
\end{eqnarray}

\begin{figure}[htb]
	\includegraphics[width=7.5cm,height=6.0cm]{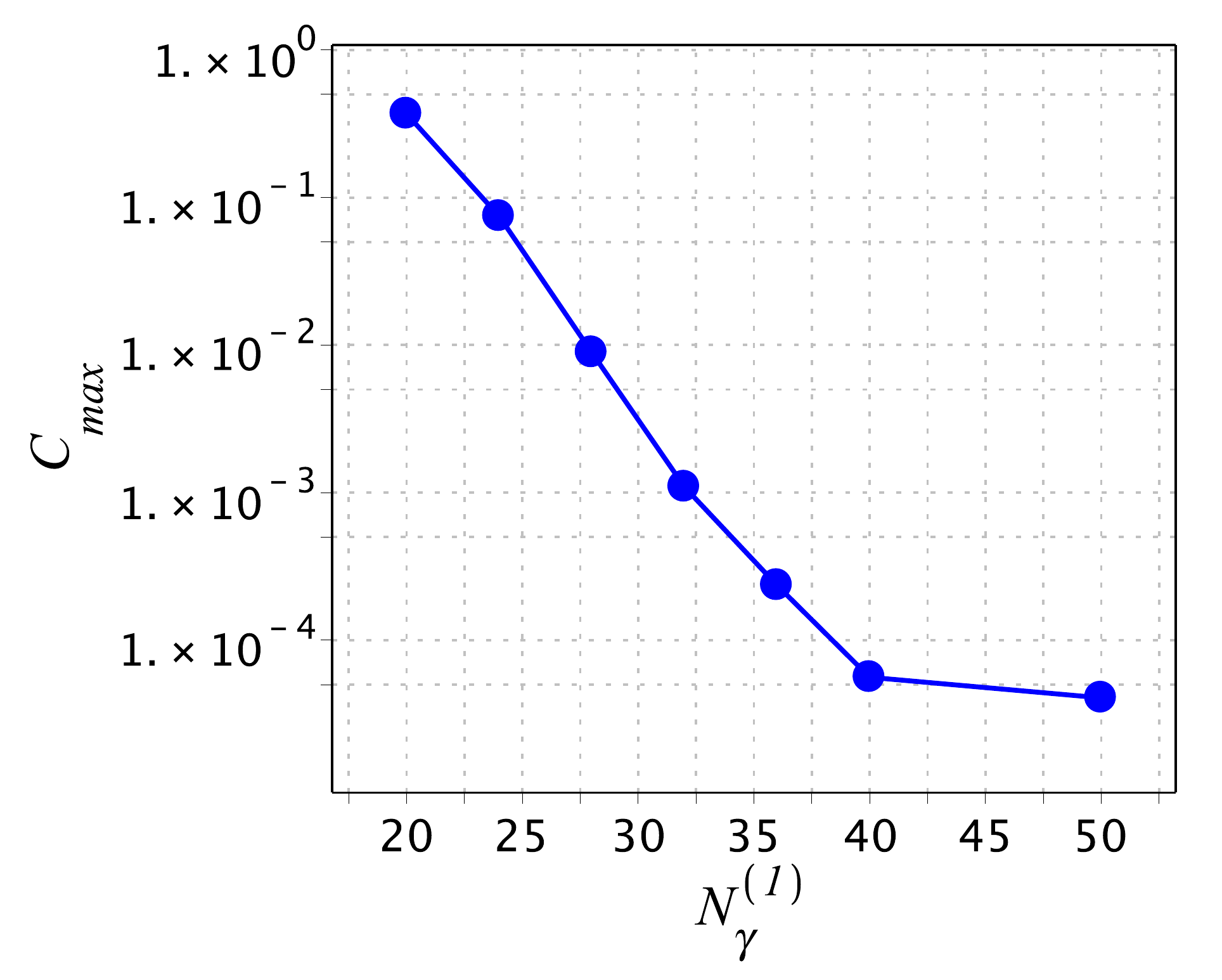}
	\caption{Exponential decay of maximum deviation $C_{\mathrm{max}}$ for the initial data (\ref{eq59}) with $r_0=3.0$, $\sigma=1$, and $A_0=0.1$. $N_\gamma^{(1)}$ is the truncation order in the first domain taken as a reference. Here $L_0=5$ and $y^{(1)}=0$ (cf. Eq. (\ref{eq19})) yielding the interface located at $r^{(1)}=L_0=5$.}
\end{figure}

\noindent where $M_B(u_0)$ is the initial Bondi mass. We have presented in Appendix the corresponding expressions for the Bondi mass, the news function, $\mathcal{N}(u,x)$, and $\omega(u,x)$. For this test, we have considered the following  initial data
\begin{eqnarray}
\bar{\gamma}_0(r,x) &=& 216 A_0 \frac{r^6 (1-x^2)^2}{(1+2r)^9}, \label{eq58} \\
\nonumber \\
\bar{\gamma}_0(r,x) &=& \frac{6 A_0 r^2 (1+r)}{(1+2r)^3}(1-x^2)\mathrm{e}^{-(r-r_0)^2/\sigma^2}, \label{eq59}
\end{eqnarray}

\noindent where $A_0$ plays the role of the initial amplitude. The first initial data is a pulse of gravitational wave more concentrated near the origin, whereas the second represents a gravitational wave Gaussian-like packet centered at $r_0$ with width $\sigma$.  %The first initial data is a pulse of gravitational wave more concentrated near the origin, whereas the second gravitational wave Gaussian-type packed centered at $r_0=3$.%, and the second data shows a pulse of gravitational more concentrated near the symmetry axis. 

To verify the convergence of the error associated with the Bondi formula, $C(u)$, we have evolved the field equations with increasing resolutions. We opted for a simple selection of the same truncation orders or the same number of collocation points in each subdomain. The radial and angular resolutions for the function $\bar{\gamma}$ in each subdomain satisfy are related by

\[N_\gamma = 2\tilde{N}_\gamma.\]

\noindent The radial and angular truncation orders of $\bar{\beta}, \bar{U}, \bar{Q}$ field functions are $N_\beta = N_U =N_Q =N_\gamma+2$ and $\tilde{N}_\beta = \tilde{N}_U =\tilde{N}_Q =\tilde{N}_\gamma+2$, respectively. For the metric function $S$, we have $N_S=N_U+2$ and $\tilde{N}_S=\tilde{N}_U+1$.

One of the considerable technical difficulties resided in calculating the Bondi mass $M_B(u)$. This task involves the determination of the mass aspect $M(u,\theta)$ (cf. Eq. (\ref{eq15})) along with the gauge terms $H, K, c$ and $L$ that appears in the asymptotic expansions of the metric functions $\beta, \gamma$, and $U$  as depicted by Eqs. (\ref{eq12}) - (\ref{eq14}). In the present numerical scheme, these quantities are obtained directly from the spectral approximations (\ref{eq24}) - (\ref{eq28}) since all basis functions satisfy the asymptotic conditions (\ref{eq12}) - (\ref{eq15}). Further, we need the same quantities list above to calculate the news function $\mathcal{N}(u,x)$. Therefore, we adapted in the code the calculations of the Bondi mass $M_B(u)$ and the news functions allowing us to obtain the deviation $C(u)$ given by Eq. (\ref{eq57}).

%We explored the error decay after fixing $y_0=-0.5$ such that the interface is now located at $r^{(1)}=L_0/3$. In Fig. 4 we present the exponential decay of $C_{\mathrm{max}}$ for the initial data (\ref{eq59}) with  amplitudes of $A_0=30$ (diamonds), $A_0=40$ (squares) and $A_0=50$ (circles). The results are better for $L_0=1.0$ (upper panel) than for $L_0=0.5$ (lower panel). 

In the Appendix, we indicated the approximations used to calculate the Bondi mass and the integrals with the news functions found in the definition of $C(u)$ (cf. Eq. (\ref{eq57})). For the initial data (\ref{eq58}), we have tested distinct values of the initial amplitude $A_0$, the map parameter, the location of interfaces. The exponential decay of the maximum deviation was observed in all cases. In Fig. 2, the red (squares) lines correspond to the initial amplitude $A_0=30$, $y^{(1)}=0$, and map parameter $L_0=0.5$ with the interface at $r^{(1)}=0.5$. The saturation of $C_{\mathrm{max}}$ occurs for $N_\gamma \geq 32$ and is about $10^{-6}\%$. In the same panel, the blue line (circles) corresponds to the following parameters: $A_0=50$, $y^{(1)}=0$, and $L_0=0.25$ producing an interface at $r^{(1)}=0.25$. Even for $N^{(1)}_\gamma = 50$ the saturation is not achieved, but the exponential decay is satisfactory. The difference in the interface locations in both cases was necessary to concentrating more collocation points near the origin for the higher amplitude. In general, $C_{\mathrm{max}}$ occurs when the Bondi mass varies more rapidly, therefore being dependend on the initial amplitude. For $A_0=30, 50$ (Fig. 2) we found $C_{\mathrm{max}} \approx 0.5$. Although not presented here, the results with domain decomposition are better than those obtained for a single domain with the same number of collocation points.

We additionally explored the error decay after fixing $y^{(1)}=-0.5$ such that the interface is now located at $r^{(1)}=L_0/3$. In Fig. 3 we present the exponential decay of $C_{\mathrm{max}}$ for the initial data (\ref{eq58}) with  amplitudes of $A_0=30$ (black circles), $A_0=40$ (blue squares) and $A_0=50$ (red diamonds). The results are better for $L_0=1.0$ (upper panel) than for $L_0=0.5$ (lower panel). 

We have set the initial amplitude $A_0=0.1$ and $\sigma=1$ for the Gaussian-type initial pulse of gravitational waves located at $r_0=3$. Although the amplitude seems small, it is enough to excite the nonlinearities of the field equations. Fig. 4 shows the log-plot of the maximum deviation of the Bondi formula, $C_{\mathrm{max}}$. It becomes clear the exponential decay until saturation of about $10^{-5}\%$ is achieved, in other words, a deviation of one part in $10^{7}$. We have integrated the approximate field equations with a fourth-order Runge-Kutta integrator. The map parameter is $L_0=5.0$ and $y^{(1)}=0$ yielding the interface located at $r^{(1)}=L_0=5.0$.  

We close this section by providing some information about the computational time-consuming in the numerical experiments. As mentioned, we have performed the numerical simulations using a fourth-order Runge-Kutta integrator with a fixed step size. 

The initial profiles (58) and (59) demand distinct choices for the map parameter and, together with the resolution or truncation orders, influence the step size selection. The initial data (58) represents a  gravitational wave packet more concentrated near the origin, which requires a smaller map parameter. On the other hand, a relatively more significant map parameter is more appropriate for a Gaussian-like wave packet (59) located at $r_0=3$.  

We indicate the value of $N_\gamma$ to characterize the other truncation orders and establish the time needed to integrate one unit of the central time, i. e. $\Delta u =1$. All numerical experiments were performed on a notebook Linux Ubuntu with processor $i7\,9750H$ and memory RAM of 32 GB. With the initial data (58), interface located at $y^{(1)}=0$, map parameter $L_0=0.5$, we have obtained: $57.28$ minutes for $N_\gamma=28$, $96.81$ minutes for $N_\gamma=32$, and $4.59$ hours for $N_\gamma=36$ with the step sizes of $1.0 \times 10^{-5}, 1.0 \times 10^{-5}$, and $0.5 \times 10^{-5}$, respectively. There is a caveat: in general the deviation reaches its maximum value for $u < 1.0$. By choosing now $y^{(1)}=-0.5$ rendering $r^{(1)}=L_0/3$ and $L_0=0.5$, we have obtained the following time costs: $119.2$ minutes for $N_\gamma=28$, $473.7$ minutes for $N_\gamma=32$, and $11$ hours for $N_\gamma=36$ with the step sizes of $5.0 \times 10^{-6}, 2.0 \times 10^{-6}$, and $2.0 \times 10^{-6}$, respectively.  

For the initial data (59) with $r_0=3$, we have set $L_0=5$ and the interface $y^{(1)}=0$. The time-costs are the following: $5.7$ minutes for $N_\gamma=28$, $19.42$ minutes for $N_\gamma=32$, $27.45$ minutes for $N_\gamma=36$, and $100$ minutes for $N_\gamma=40$. The step sizes are: $1.0 \times 10^{-4}$, $5.0 \times 10^{-5}$, $5.0 \times 10^{-5}$, and $2.0 \times 10^{-5}$, respectively.

\begin{figure}[htb]
	\includegraphics[width=6.5cm,height=6.5cm]{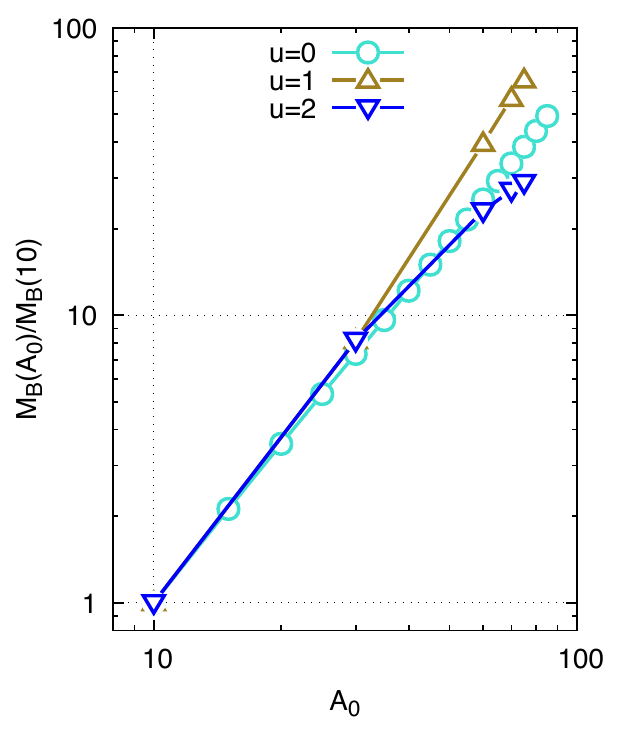}
	\includegraphics[width=6.5cm,height=6.5cm]{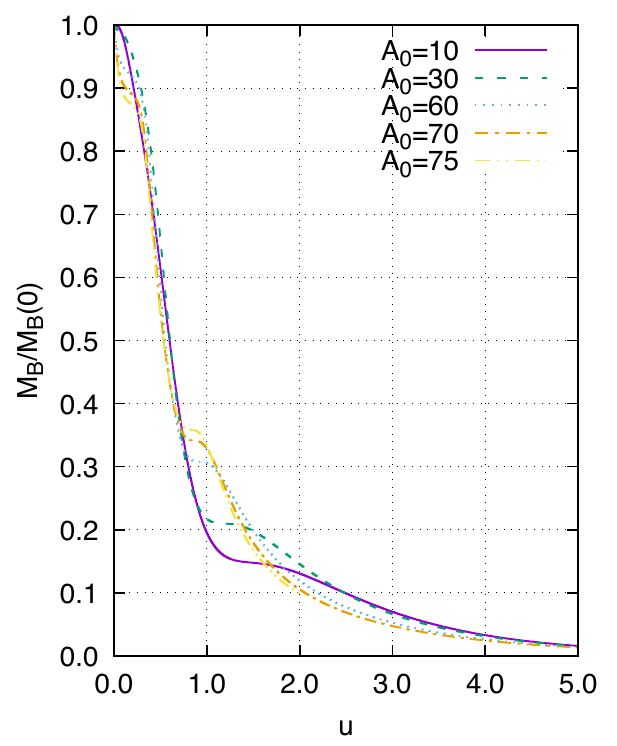}
	\caption{Quadratic dependence of the Bondi mass with the initial amplitude for $u=0,\, 1, \,2$ (top plot). For each time the Bondi mass is normalized with the Bondi mass for the smallest considered amplitude ($A_0=10$). Bondi mass normalized by the initial Bondi mass for each amplitude as a function of time (bottom plot). We have used $N_\gamma^{(1)}=30$ for $A_0=10, \,30$ and $N_\gamma^{(1)}=40$ for other amplitudes.} 
	%\caption{Decay of the Bondi masses for $A_0=10,30$ in the first panel, and $A_0=60,70$ and $75$ in the second panel. The initial data is given by (\ref{eq58}) We have used $N^{(1)}_\gamma=30, 40$, respectively for the first and second panels.}
\end{figure}

\begin{figure}[htb]
	\includegraphics[width=6.5cm,height=5.cm]{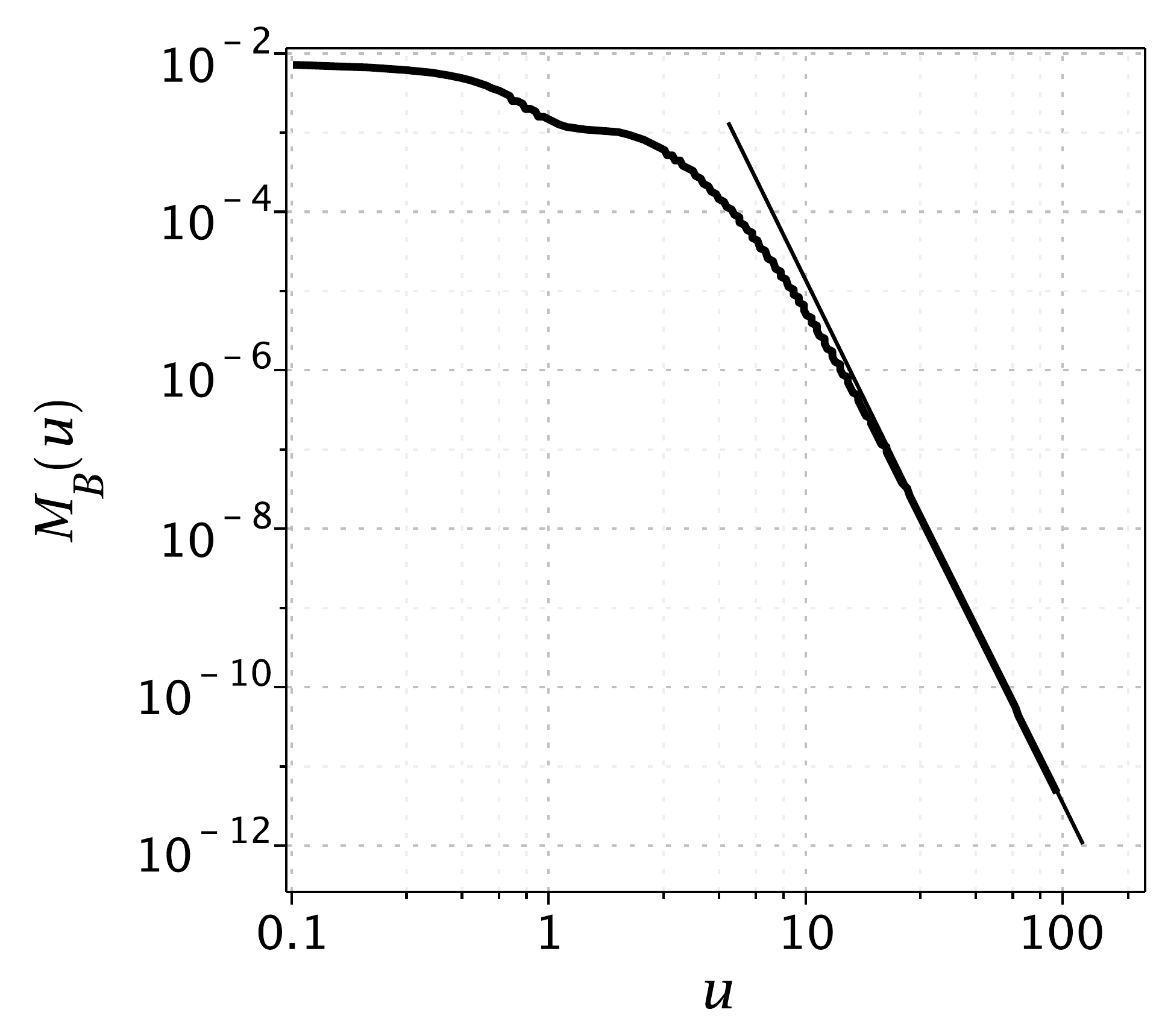}
	\caption{Power-law late time decay of the Bondi masses for $A_0=10$ for the initial data given by (\ref{eq58}). Here $M_B(u) \sim u^{-6.59}$.}
\end{figure}

%%%%%%%%%%%%%%%%%%%%%%%%%%%%%%%%%%%%%%%%%%%%%%%%%%%%%%%%%%%%%%%%%%
\section{The Bondi mass decay and the gravitational wave patterns}

As a physical application, we generate the gravitational waveforms emitted by an initially ingoing pulse towards the origin. It is well established that asymptotic flat spacetimes admitting gravitational waves have a nonvanishing term of order $r^{-1}$ in the Weyl tensor when viewed far from the source \cite{newman_penrose}. It means, according to the Peeling Theorem \cite{sachs}, that the curvature tensor has the same algebraic structure as a plane wave.  Consequently, we are interested in the term of the Weyl scalar $\Psi_4$ that falls as $r^{-1}$ at the future null infinity. 

In the original Bondi frame \cite{bondi} characterized by $H=L=K=0$, one can show, after choosing an appropriate null tetrad basis (see Appendix), that 
\begin{equation}
\Psi_4(u,r,\theta) \simeq -\frac{c(u,\theta)_{,uu}}{r},  \label{eq60}
\end{equation}

\noindent where $c(u,\theta) = \lim_{r \rightarrow \infty}\,(r \gamma)$ (cf. Eq. (\ref{eq12})). In the frame we are adopting, the Weyl scalar has a similar structure:% However, in the frame we are adopting, the Weyl scalar has the  following asymptotic structure:
\begin{eqnarray}
%\Psi_4(u,r,\theta) = \Psi^{(0)}_4(u,\theta)+ \frac{\Psi^{(1)}_4(u,\theta)}{r} +  \mathcal{O}(r^{-2}), \label{eq61} 
\Psi_4(u,r,\theta) = \frac{\Psi^{(1)}_4(u,\theta)}{r} +  \mathcal{O}(r^{-2}), \label{eq61} 
\end{eqnarray}

%\noindent Here, $\Psi^{(0)}_4(u,\theta)$ is expressed in terms of the gauge functions $H,K,L,$ and their derivatives with respect to $u$ and $\theta$.

\noindent with the term $\Psi^{(1)}_4(u,\theta)$ describing the waveforms perceive by a distant observer.
%We are interested in the term $\Psi^{(1)}_4(u,\theta)$ to describe correctly the wave patterns received by a distant observer. 
The full expression of  $\Psi^{(1)}_4(u,\theta)$ is quite large \cite{bishop_rezolla}, but, for the sake of convenience, we exhibit this function evaluated at $\theta=\pi/2$ or $x=0$. It reads 
%
%\begin{widetext}
{\small
\begin{eqnarray}
	\Psi^{(1)}_4&&(u,x=0) =\bigg\{ \mathrm{e}^{2(\bar{H}-\bar{K})}\bigg[\left(\bar{H}_{,xx}-\frac{1}{2}\bar{K}_{,xx}+\bar{K}-4 \bar{H}+1\right) \times \nonumber \\
	\nonumber \\
	&&\bar{L}_{,x}+\left(\bar{K}_{,xx}+2\bar{H}_{,xx}-2\bar{K}-8 \bar{H}+1\right)\,\bar{K}_{,u} -\frac{1}{4}\bar{L}_{,x,x,x} \nonumber \\
	\nonumber \\
	&&+4\bar{H}_{,u}-\bar{H}_{,uxx}\bigg] - \bar{H}_{,u}\bar{L}_{,x}\bar{c} -\frac{1}{2}\bar{L}_{,x}^2\bar{c}+\frac{3}{2}\bar{L}_{,x}\bar{c}_{,u} \nonumber \\
	\nonumber \\
	&&+\frac{1}{2}\bar{L}_{,ux}\bar{c}+2\bar{H}_{,u}\bar{c}_{,u}-\bar{c}_{,uu}\bigg\}_{x=0}, \label{eq62}
\end{eqnarray}}
%\end{widetext}

\noindent where we have taken into account the symmetries of the metric functions with respect to $x=0$, $H=(1-x^2)^2\bar{H}$, $K=(1-x^2)\bar{K}$, $L=\sqrt{1-x^2}\bar{L}$ and $c=(1-x^2)\bar{c}$.  As expected, we recover the original expression (\ref{eq60}) of $\Psi_4$ in the Bondi frame.

\begin{figure}[htb]
%	\includegraphics[width=6.3cm,height=5.cm]{fig7a}
 	%\hspace{0.7cm} \includegraphics[width=6.2cm,height=7.5cm]{psi4.pdf}
	%\hspace{-3.8cm}\includegraphics[width=13cm,height=5.5cm]{psi4_new}
  %\center{\includegraphics[scale=0.27]{psi4_new}}
\center{\hspace{1cm}\includegraphics[width=7.5cm,height=5.5cm]{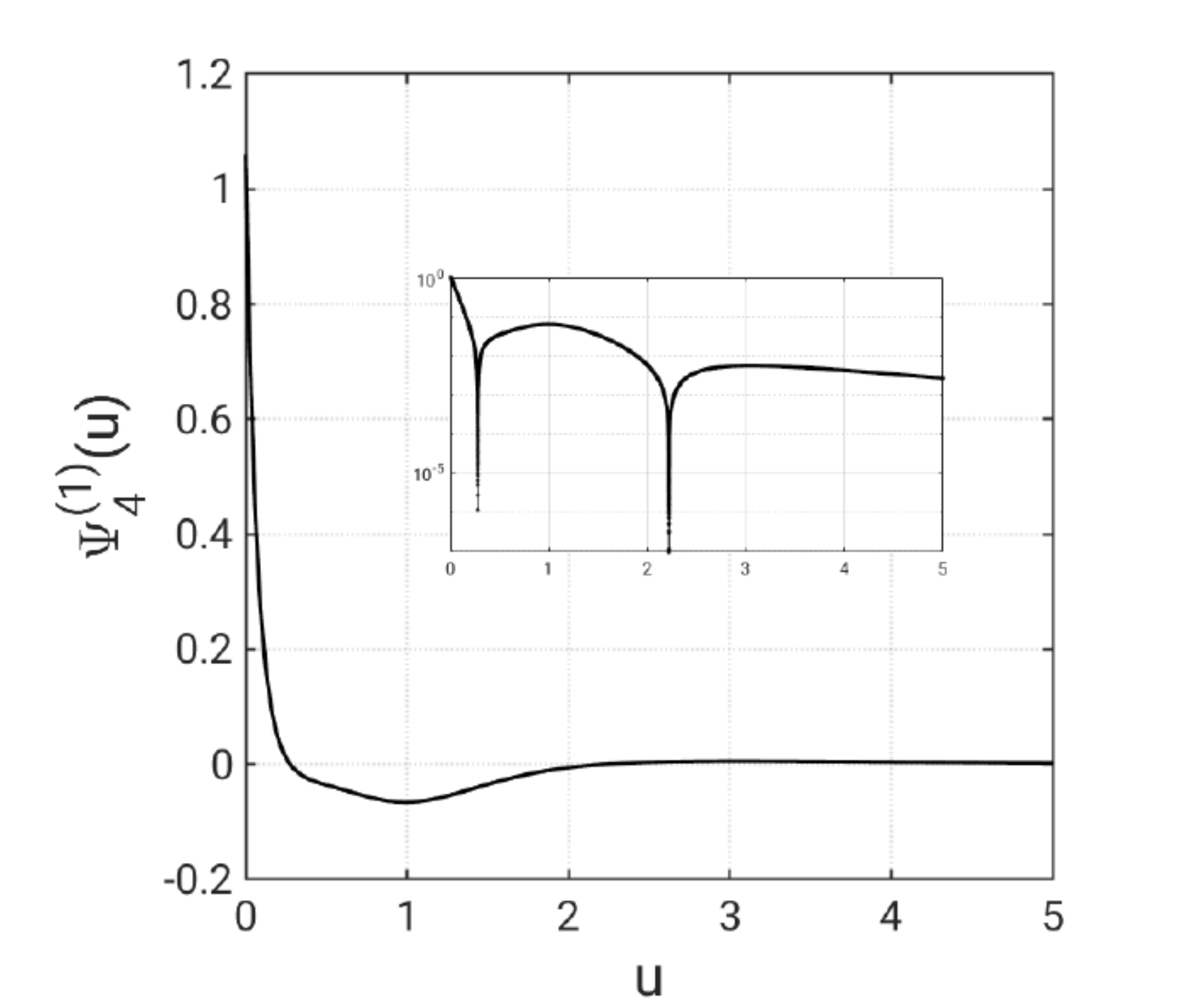}
	\includegraphics[width=7.5cm,height=5.cm]{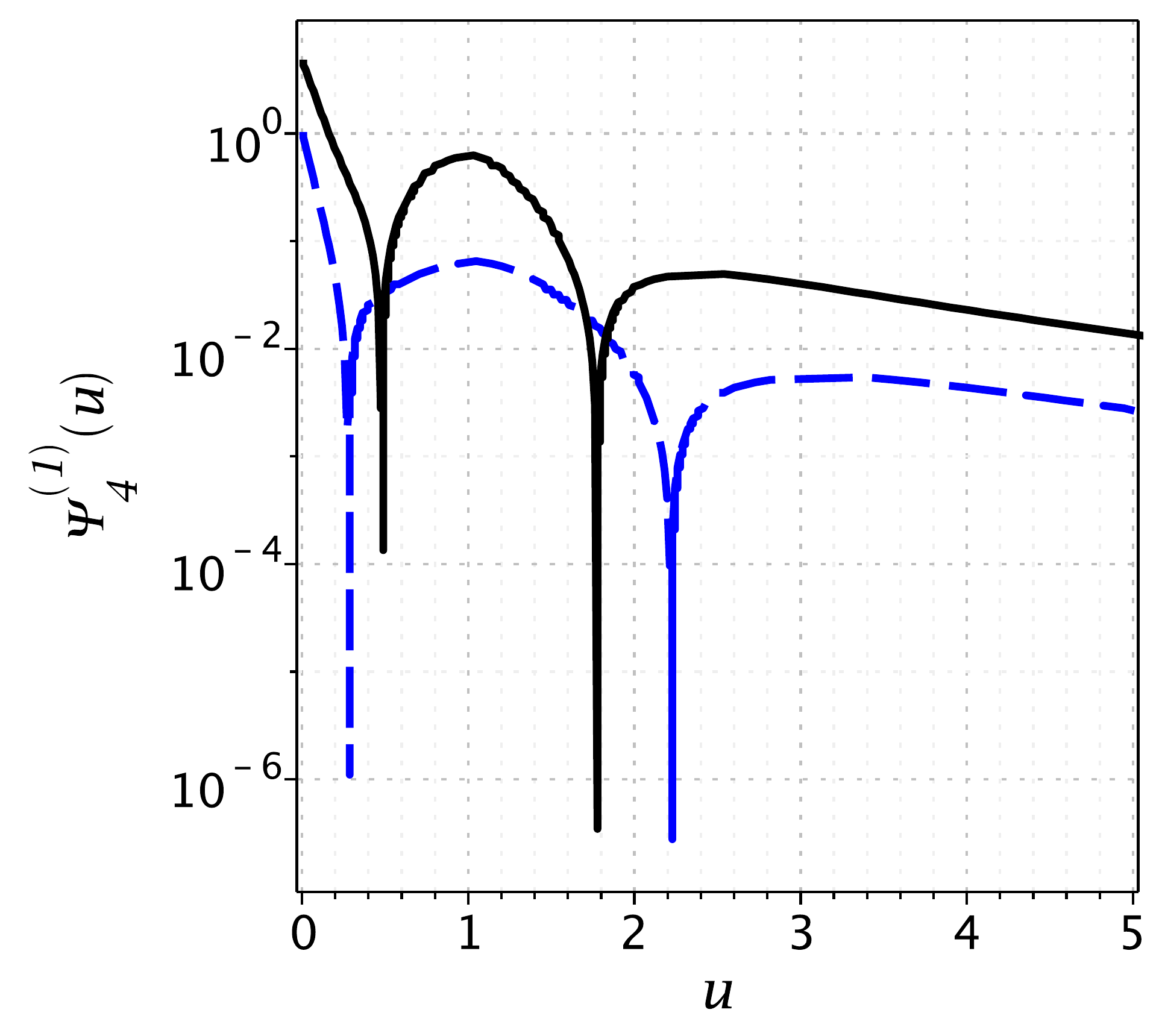}}
	\caption{Upper panel: plot of $\Psi_4^{(1)}(u)$ together with the inset showing the corresponding logplot. The amplitude is $A_0=10$. Lower panel: logplot of $\Psi_4^{(1)}(u)$ for $A_0=10$ (dashed line) and $A_0=30$ continuous line.}
\end{figure}

In the sequence, we present the wave patterns generated by different initial incoming pulses described by $\bar{\gamma}(u_0,r,x)=\bar{\gamma}_0(r,x)$. We first considered the initial data Eq. (\ref{eq58}) previously used for the code validation. In that case, the pulse is close to the symmetry axis, meaning that it bounces after hitting the symmetry axis in a short time interval. As a consequence, gravitational waves carry away most of the mass. 

Figure 5 (top plot) displays the expected quadratic dependence of the Bondi mass with the initial amplitude for $u=0, 1, 2$. For each time the Bondi mass is normalized with the Bondi mass for the smallest considered amplitude ($A_0=10$). Clearly, with evolution the initial linear dependence (in the log-log plot of the Bondi mass) deviates. This is a measure of the transient non-linear effects. In the sequence, in the bottom plot we display the Bondi mass normalized by the initial Bondi mass for each amplitude (whose dependence with the initial amplitude is shown in the top plot) as a function of time. Certainly, in a different way, the bottom plot displays the non-linear effects for different initial amplitudes and under the action of time.
The Bondi mass decay corresponding to the initial amplitudes $A_0=10,\,30,\,60,\,70$ and $75$. In all cases, the gravitational waves extract about more than $50\%$ of the initial mass when $u \approx 1.0$. The relative amount of mass loss is greater for smaller initial amplitudes; more precisely, we have found that for $A_0=10, 30,60,70,75$, about $80.8\%, 78.3\%,69.4\%,67.15\%, 64.28\%$ respectively, of the initial mass is radiated away at $u \approx 1.0$. For higher amplitudes, $A_0=60,70,75$, there is a fast initial mass decay as soon as the pulse moves towards the symmetry axis. It seems to us that the behavior of $\dot{M}_B\ne 0$ at $u=0$ is related with the backscattering and consequently with the initial flux of energy at infinity; 
the backscattering is stronger for the higher initial amplitude. One can think it is a non linear effect, but a more careful study is required because a previous result indicates that could be present even in the linear regime. In this sense,  
we note that the same behavior appears in the setting studied in Ref. \cite{b14} for the Einstein-Klein-Gordon system, using ingoing null cones. In this work, which assumes the same symmetry, $\dot{M}_B \ne 0$ at $u=0$ is stronger with the increasing of the angular structure (in the linear regime).

We have also observed (see Fig. 6) the late time power-law decay of the Bondi mass, $M_B(u) \sim u^{-6.59}$. We have set $A_0=10$, but the same law is obtained for greater initial Bondi masses.  
\begin{figure}[htb]
    \includegraphics[width=6.5cm,height=5.cm]{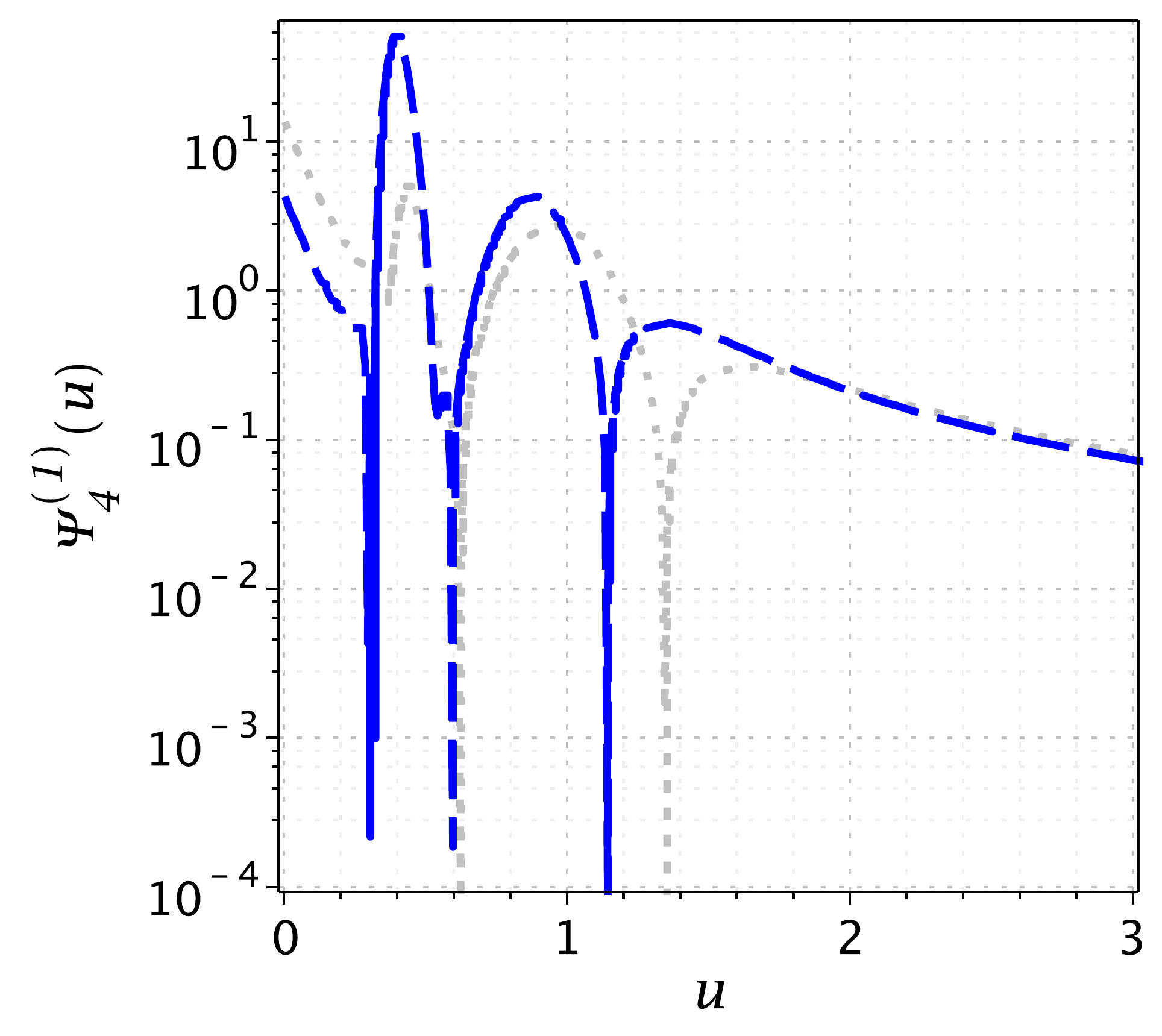}\\%{psi4_60_70_id1_3}\\
	\hspace{1.4cm} (a)\\
	\includegraphics[width=6.5cm,height=5.cm]{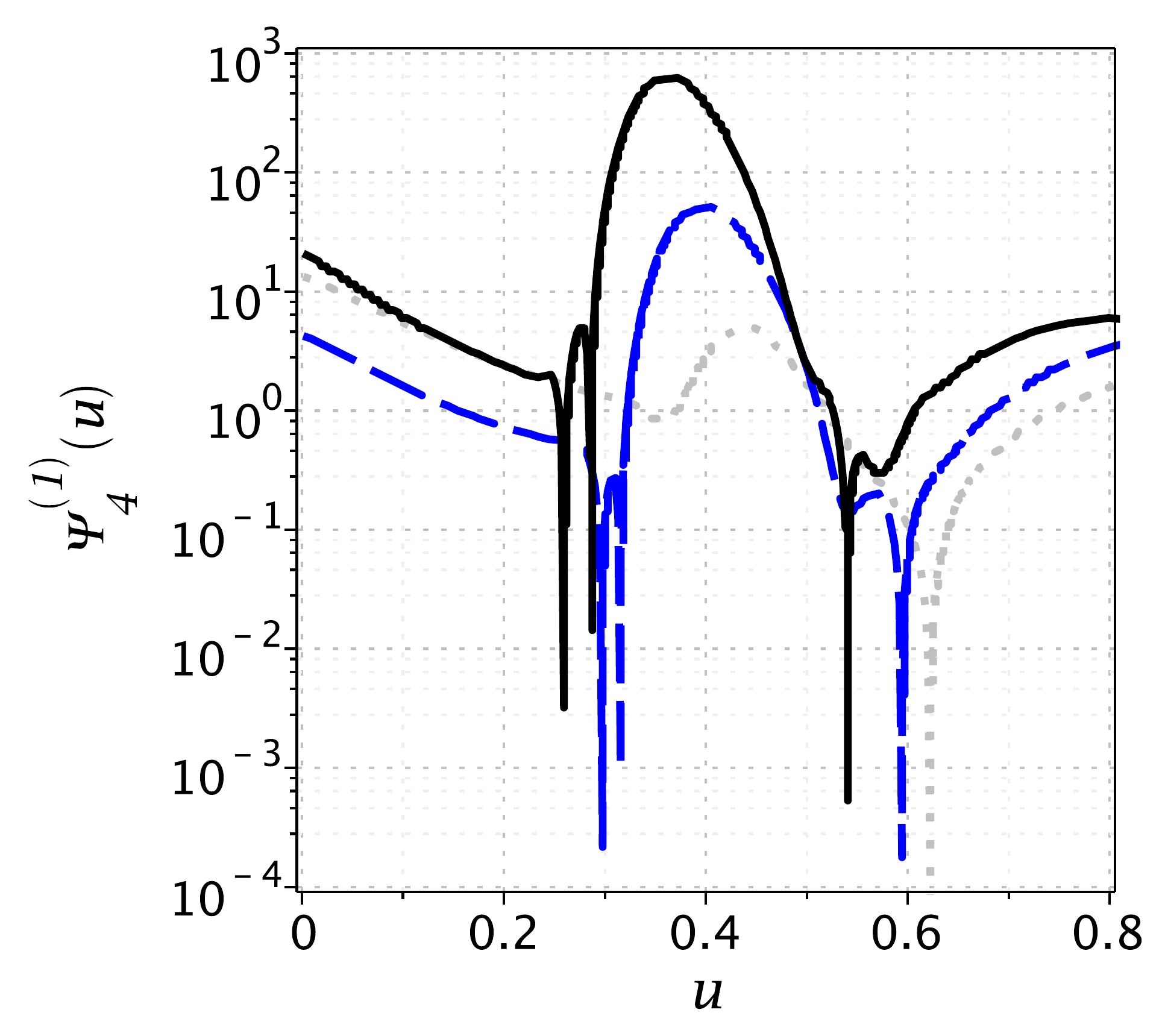}\\%{psi4_60_70_75_id1_zoom_2}\\
	\hspace{1.4cm} (b)
	\caption{(a) Log-linear plots of $\Psi_4^{(1)}(u)$ for $A_0=60,70$ represented by dashed and dotted lines, respectively. (b) Zoom of the region $0 \leq u \leq 0.8$ to visualize the appearence of small oscillations. The continuous line corresponds to $A_0=75$.}
\end{figure}

We expect that the nuances of the mass decay leave imprints in the observed pattern of the gravitational waves described by Eq. (\ref{eq62}).  Fig. 7 (upper panel) shows the plots of $\Psi_4^{(1)}(u)$ corresponding to $A_0=10$ together with its log-plot in the inset.  There is a rapid variation $\Psi_4^{(1)}(u)$ in the interval $0 \leq u \leq 1$ and an oscillation. We can better visualize the effect of increasing the initial amplitude from $A_0=10 $ to $A_0=30$ on the wave pattern with the log-plot of $\Psi_4^{(1)}(u)$ as shown in Fig. 7 (lower panel). We observe that the original oscillation takes place in a shorter time interval, meaning increasing the frequency. 

Furthermore, a more significant change occurs in the highly nonlinear regime in which $A_0=60, 70$ and $75$. The Bondi mass decay is similar in all cases, but increasing the initial amplitude produces more oscillations in shorter time intervals, as shown in Fig. 8. The dashed and dotted lines in Fig. 8(a) correspond to $A0=60$ and $70$, respectively. In the zoom displayed in Fig. 8(b), with the plots of $\Psi_4^{(1)}(u)$ for $A_0=60,70$ and $75$ (continuous line).  

\begin{figure}[htb]
	\includegraphics[width=6.5cm,height=5.cm]{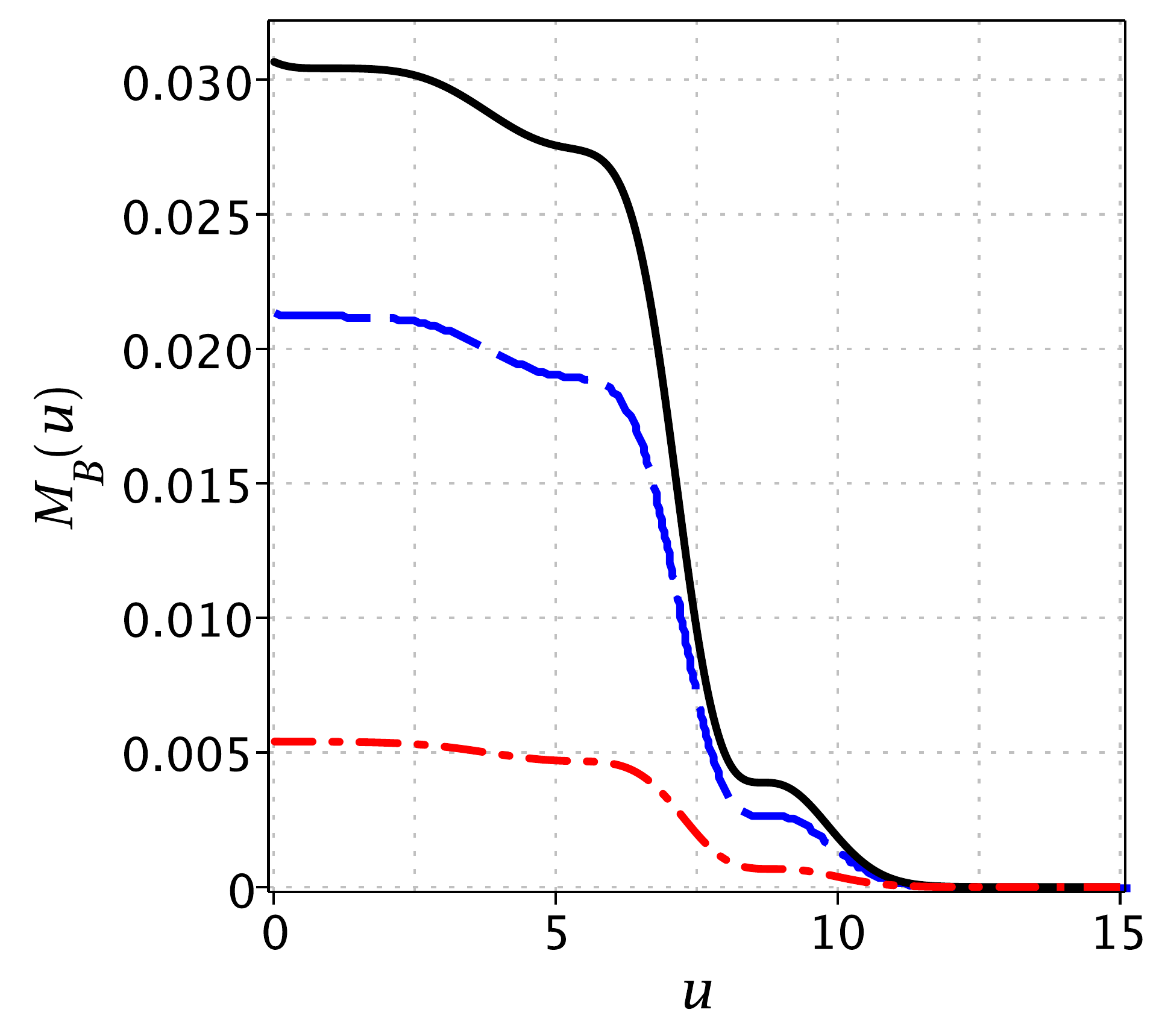}\\%{MB_0_007_0_10_0_12_id3}\\
	\hspace{1.4cm} (a)\\
	\includegraphics[width=6.5cm,height=5.cm]{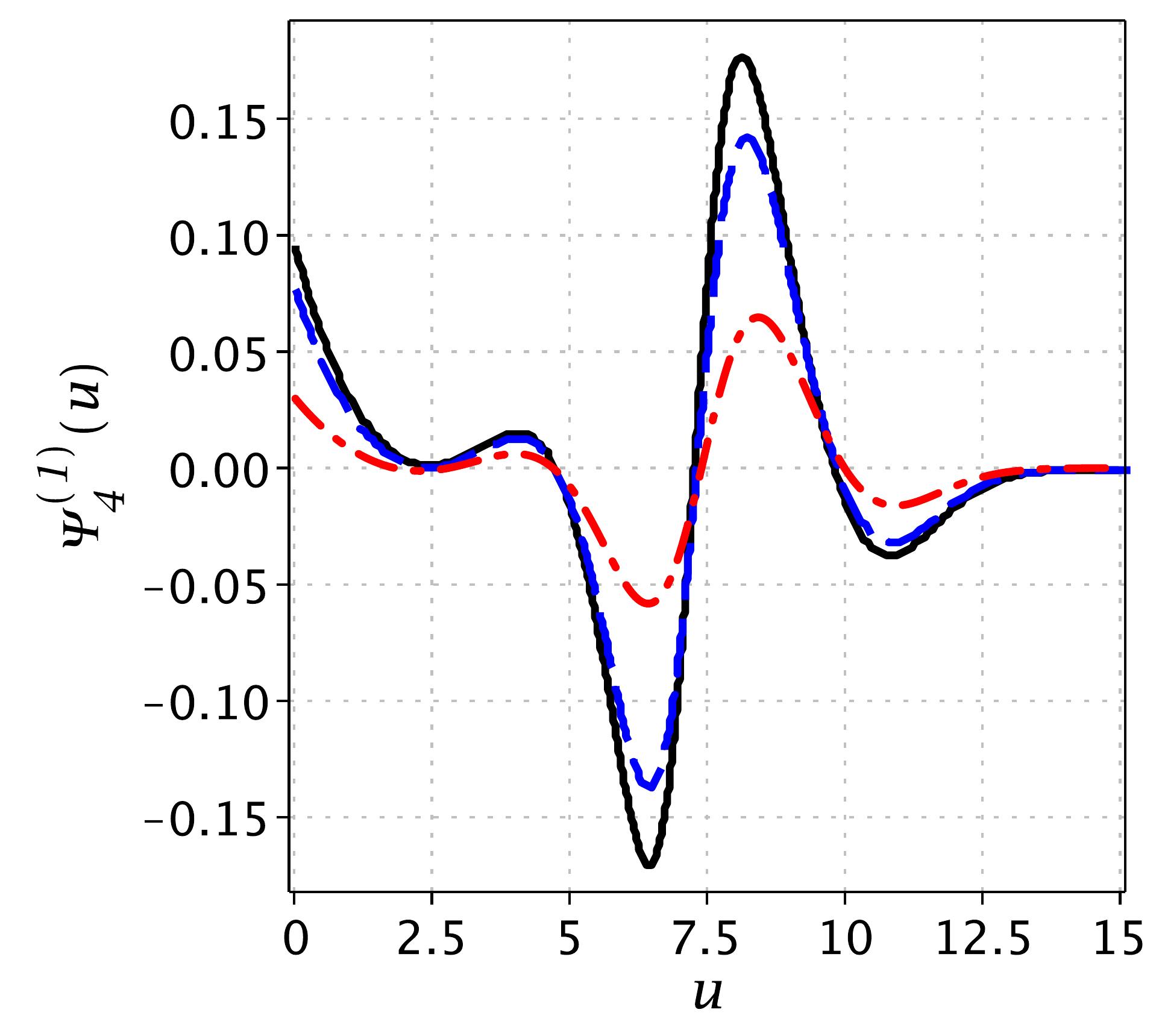}\\%{psi4_0_007_0_10_0_12_id3}\\
	\hspace{1.4cm} (b)\\ 
	\caption{(a) Decay of the Bondi mass for the initial data (\ref{eq59}) with $A_0=0.05$ (dashed-dotted line), $A_0=0.10$ (dashed line) and $A_0=0.12$ (continuous line). (b) Following the same convention, we show the behavior of $\Psi_4^{(1)}(u)$ for $A_0=0.05,0.10$ and $A_0=0.12$.}
\end{figure}

\begin{figure}[htb]
	\includegraphics[width=6.5cm,height=5.cm]{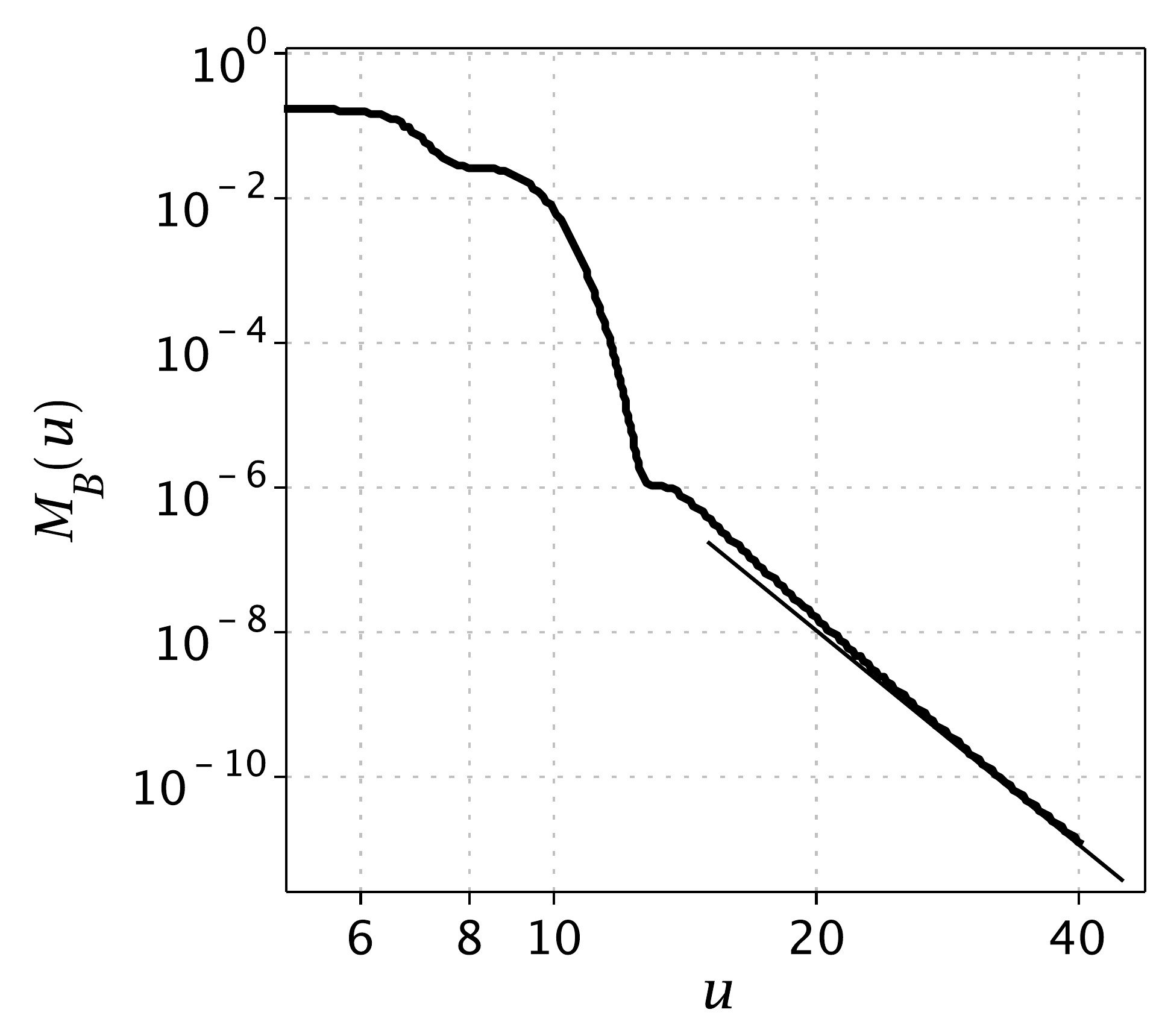}%{MBondi_A0_0_3_id3_power_law}
	\caption{Power-law late time decay of the Bondi mass for  $A_0=0.3$ for the initial data given by (\ref{eq59}). Here $M_B(u) \sim u^{-9.83}$.}
\end{figure}

We have now considered the initial data (\ref{eq59}), but with $r_0=4$ and $\sigma=1$. The simulations have taken a long time necessary for the pulse to hit the symmetry axis and bounce away. In Fig. 9(a), we present the decay of the Bondi mass for $A_0=0.05,0.10$ and $A_0=0.12$, where the nonlinear regime progressively enters into action. Notice the substantial increase in the initial Bondi mass when the amplitude $A_0$ changes. From $A_0=0.05$ to $A_0=0.10$, the initial Bondi mass increases almost four times, and from $A_0=0.10$ to $A_0=0.12$, the initial Bondi mass increases approximately $50\%$. Since the initial pulse is not close to the symmetry axis, there is no considerable mass loss in the first moments; only starting at $u \approx 6$ there is a huge mass extraction by gravitational radiation, mainly for $A_0=0.10$ and $A_0=0.12$. The resulting patterns of $\Psi_4^{(1)}(u)$ depicted in Fig. 9(b) are sensitive to the decay of the Bondi mass corresponding to the initial data under consideration. The considerable variation of the Bondi mass left its imprint producing the oscillatory pattern $\Psi_4^{(1)}(u)$. Finally, in Fig. 10, we present the power-law behavior of the Bondi mass at late times, where we have found $u \sim u^{-9.83}$ which is independent of the initial amplitude.

\section{Final comments}

We have implemented a simple domain decomposition spectral code based on the Galerkin-Collocation method applied to the field equations of the celebrated Bondi problem. The present task is a part of a program of implementing domain decomposition in problems of interest in numerical relativity \cite{alcoforado_critical,alcoforado_cauchy,barreto_DD,barreto_cylindrical_DD}. We have presented with detail the following topics:

\begin{enumerate}
	
	\item[(i)]  The new computational scheme of establishing the non-overlapping subdomains after the compactification of the spatial domains (cf. Fig. 1).
	
	\item[(ii)] The definition of radial and angular basis functions for the metric functions $\gamma,\,\beta,\,U$, and $V$.
	
	\item[(iii)] The use of two sets of collocation points in each subdomain connected with the functions $\beta,\,U$ and $V$. 
	
	\item[(iv)] The incorporation of the transmission conditions into the field equations. In particular, we have considered the most simple form (see Refs. \cite{alcoforado_critical,alcoforado_cauchy} ) for the evolution equation, since for hyperbolical problems, there is no unique form for the transmission conditions.
	
	\item[(v)] The approximations of the hypersurface equations as systems of algebraic equations and the evolution equation as a system of ordinary differential equations for the modes associated with the metric function $\gamma$. 
	
\end{enumerate}

The combination of features of both Galerkin and Collocation methods rendered a stable and exponentially convergent code which is attested with the verification of the Bondi formula. We have proceeded with the decay of the Bondi mass for high amplitude initial data, and we have exhibited its late time power-law decay. Since the decay of the Bondi mass is connected with the gravitational wave extraction, we have shown the waveforms at the future null infinity. We obtain the wave-forms from the Weyl scalar $\Psi_4$, which we have calculated in the present gauge, that is, the one characterized by non-vanishing asymptotic quantities $H,\, K$, and $L$. 

%We indicate some directions of the present work that are under investigation. The first is to study the collapse of gravitational waves more thoroughly with or without a matter content, such as a massless scalar field. In this case, the search for the apparent horizon formation is the main target and the exploration of the critical collapse. Another possibility is the inclusion of more subdomains that will produce an increase in code accuracy.

We indicate some directions of the present work that are under investigation. The first is to study the collapse of gravitational waves more thoroughly with or without a matter content, such as a massless scalar field. In this case, examining the critical behavior in axisymmetric collapse is of great interest. Several works on the collapse of Brill waves have touched this issue without a complete understanding \cite{choptuik_03,hilditch_13,ledvinka_21}.
It is possible to include more subdomains to increase the code resolution for a better description of the implosion of gravitational waves.

%\acknowledgments
\section*{Acknowledgments}

W. B. acknowledges the financial support by Universidade do Estado do Rio de Janeiro (UERJ) and Funda\c{c}\~ ao\ Carlos Chagas Filho de Amparo a Pesquisa do Estado do Rio de Janeiro (FAPERJ); also thanks the hospitality of the Departamento de F\'isica Te\'orica at Instituto de F\'isica (UERJ). H. P. O. acknowledges the financial support of the Brazilian Agency CNPq and the support of FAPERJ. This study was financed in part by the Coordena\c c\~ao de Aperfei\c coamento de Pessoal de N\'ivel Superior - Brasil (CAPES) - Finance Code 001.

\appendix

\section{Bondi mass and the news function}

The Bondi mass in the present gauge \cite{isaacson} is given by
%
%\begin{widetext}
\begin{eqnarray}
M_B(u) &&= \int_{-1}^1\,\frac{\mathrm{e}^{-2K}}{\omega}\bigg[\frac{\mathrm{e}^{2K}}{2}M + \frac{1}{4} (1-x^2) c^{\prime\prime} -x c^{\prime} -\frac{1}{2}c \times\nonumber \\
\nonumber \\
&&(1-x^2)c^\prime(H^\prime+K^\prime) - (1-x^2)c\bigg(H^{\prime2}-2H^\prime K^\prime-\nonumber \\
\nonumber \\
&& -K^{\prime2}+\frac{1}{2}H^{\prime\prime}+\frac{1}{2}K^{\prime\prime}\bigg) + 2 x c (H^\prime+K^\prime)\bigg]dx,
%\nonumber \\
\end{eqnarray}
%\end{widetext}

\noindent where 

\[\omega(u,x)=\frac{2\, \mathrm{e}^{2 K}}{(1+x)\mathrm{e}^{\Delta}+(1-x)\mathrm{e}^{-\Delta}},\]

\noindent and

\[\Delta(u,x)=\int_0^x\,\frac{(\mathrm{e}^{2 K}-1)}{1-x^2}\,dx.\]

\noindent Here prime means derivative with respect to $x=\cos \theta$. The news function can be written as 
\begin{eqnarray}
\mathcal{N} &&= \mathrm{e}^{2H}c_{,u} - \frac{\mathrm{e}^{-2H}}{2c} (c^2 L)^\prime  + \frac{1}{2} (1-x^2) \mathrm{e}^{-2K} +
\nonumber \\
&&+\frac{1}{2}(1-x^2)\mathrm{e}^{-2K}\bigg(2H^{\prime\prime}+4H^{\prime2}+ \frac{\omega^{\prime\prime}}{\omega}-2\frac{\omega^{\prime2}}{\omega^2}\bigg).
%\nonumber \\
\end{eqnarray}

\noindent We have calculated the integral (A1) using Gauss-Legendre quadrature formulae schematically as
\begin{equation}
M_B(u) \approx \sum_{k=0}^{N_{q}}\,\left(...\right)_k wq_k,
\end{equation}

\noindent where $\left(...\right)_k$ means the integrand evaluated at the quadrature collocation points, $N_q$ indicates the number of quadrature points, and $wq_k$ represents the weights. Similarly, we calculate the integral involving the news function (RHS of Eq. (\ref{eq16})), and the integration with respect to $u$ (cf. Eq. \ref{eq57}) is performed using the Newton-Cotes formula.

\section{Tetrad basis}

We have used the following tetrad basis for the calculation of the Weyl scalar $\Psi_4$ \cite{nerozzi}:

\begin{eqnarray}
l^\mu &=& \left(0,-\mathrm{e}^{-2 \beta},0,0\right) \\
\nonumber \\
n^\mu &=& \left(-1,\frac{V}{2r}-\frac{1}{2}U^2 r^2\mathrm{e}^{-2(\gamma- \beta)},0,0\right) \\
\nonumber \\
m_\mu &=& \left(0,-\frac{1}{\sqrt{2}} U\mathrm{e}^{\gamma-2\beta},\frac{1}{\sqrt{2}r} \mathrm{e}^{-\gamma},\frac{i}{\sqrt{2}r \sin \theta} \mathrm{e}^{\gamma}\right)
\end{eqnarray}

\end{document}